\documentclass{emulateapj}

\usepackage[usenames]{color}
\usepackage{apjfonts}

\begin{document}

\title{Total infrared luminosity estimation of resolved and unresolved galaxies}

\author{M. Boquien\altaffilmark{1}, G. Bendo\altaffilmark{2}, D. Calzetti\altaffilmark{1}, D. Dale\altaffilmark{3}, C. Engelbracht\altaffilmark{4}, R. Kennicutt\altaffilmark{5}, J. C. Lee\altaffilmark{6,7}, L. van Zee\altaffilmark{8}, J. Moustakas\altaffilmark{9,10}}
\email{boquien@astro.umass.edu}

\altaffiltext{1}{University of Massachusetts, Department of Astronomy, LGRT-B 619E, Amherst, MA 01003, USA}
\altaffiltext{2}{Astrophysics Group, Imperial College, Blackett Laboratory, Prince Consort Road, London SW7 2AZ, UK}
\altaffiltext{3}{Department of Physics and Astronomy, University of Wyoming, Laramie, WY 82071, USA}
\altaffiltext{4}{Steward Observatory, University of Arizona, Tucson, AZ 85721, USA}
\altaffiltext{5}{Institute of Astronomy, University of Cambridge, Madingley Road, Cambridge CB3 0HA, UK}
\altaffiltext{6}{Carnegie Observatories, 813 Santa Barbara Street, Pasadena, CA 91101, USA}
\altaffiltext{7}{Hubble Fellow}
\altaffiltext{8}{Astronomy Department, Indiana University, 727 East 3rd Street, Bloomington, IN 47405, USA}
\altaffiltext{9}{Center for Cosmology and Particle Physics, 4 Washington Place, New York University, New York, NY 10003, USA}
\altaffiltext{10}{Center for Astrophysics and Space Sciences, University of California, San Diego, 9500 Gilman Drive, La Jolla, CA 92093, USA}

\begin{abstract}
The total infrared (TIR) luminosity from galaxies can be used to examine both star formation and dust physics. We provide here new relations to estimate the TIR luminosity from various {\em Spitzer} bands, in particular from the 8~$\mu$m and 24~$\mu$m bands. To do so, we use 45\arcsec\ subregions within a subsample of nearby face-on spiral galaxies from the Spitzer Infrared Nearby Galaxies Survey (SINGS) that have known oxygen abundances as well as integrated galaxy data from the SINGS, the Local Volume Legacy Survey (LVL) and \cite{engelbracht2008a} samples. Taking into account the oxygen abundances of the subregions, the star formation rate intensity, and the relative emission of the polycyclic aromatic hydrocarbons at 8~$\mu$m, the warm dust at 24~$\mu$m and the cold dust at 70~$\mu$m and 160~$\mu$m we derive new relations to estimate the TIR luminosity from just one or two of the {\em Spitzer} bands. We also show that the metallicity and the star formation intensity must be taken into account when estimating the TIR luminosity from two wave bands, especially when data longward of 24~$\mu$m are not available.
\end{abstract}

\keywords{infrared: galaxies}

\section{Introduction}

With the end of the {\em Spitzer} cold phase and the widespread availability of 8~$\mu$m and 24~$\mu$m bands observations in the archives, the availability of relations to determine the total infrared (TIR) emission from these wave bands by themselves is crucial to efficiently exploit the archives. In addition, the {\em Herschel} Space Observatory observes dust emission at rest-frame 24~$\mu$m and longer wavelengths for galaxies redshifted to $z\simeq1.5$ in the PACS 60~$\mu$m and longward bands with a spatial resolution as good as the {\em Spitzer} 24~$\mu$m band, for instance. Thus, estimating the total infrared flux from these bands becomes crucial for measuring total infrared fluxes using {\em Herschel} data. Finally, the advent of new instrumentation in the coming years, such as the James Webb Space Telescope (JWST) or the Atacama Large Millimeter Array (ALMA) will also open a new window on the infrared emission of nearby and distant galaxies. For instance, the 18~$\mu$m JWST/MIRI band and the 350~$\mu$m ALMA band will probe the rest-frame 8~$\mu$m and the 160~$\mu$m emission of $z\simeq1.2$ galaxies.

Determining the total infrared emission using the 8-160~$\mu$m {\em Spitzer} bands with equation 4 from \cite{dale2002a} or equation 22 from \cite{draine2007a} yields a better estimate of the total infrared flux than using a single wave band as a proxy for the total infrared flux, but they necessitate 3 and 4 infrared bands respectively. Indeed, these relations necessitate using the much lower resolution 70~$\mu$m and 160~$\mu$m data, which have resolutions of 18\arcsec\ and 40\arcsec, respectively. The poorer resolution of the 70~$\mu$m and 160~$\mu$m bands strongly constrains the scales on which total infrared fluxes can be measured, even for local galaxies. Attempts to derive a relation to estimate the total infrared emission from the 8~$\mu$m and 24~$\mu$m bands have been made by \cite{calzetti2005a} using NGC~5194 (M51) data, \cite{perez2006a} using NGC~3031 (M81) data, and \cite{thilker2007a} using M33 data. For each galaxy, the scatter around the relation is about 40\%. \cite{rieke2009a} recently showed that using the 24~$\mu$m band only provided good results. Using the 24~$\mu$m and the 70~$\mu$m bands \cite{papovich2002a} have also provided an estimate of the TIR emission with a similar uncertainty. However, these relations may be applicable only to galaxies with similar metallicities and star formation rate intensities (Calzetti et al. 2010, submitted; Li et al. 2010, in preparation). Metallicity variations have been associated with variations in mid- and far-infrared colors, as has been observed in metal-poor galaxies \citep[e.g.][]{engelbracht2005a, engelbracht2008a}. Indeed, the \cite{calzetti2005a} relation, which was derived using data from a very metal-rich galaxy, underestimates the total infrared emission by a factor of a few in metal-poor dwarf galaxies \citep{cannon2005a,cannon2006a,cannon2006b}.

In this article we derive relations to estimate the total infrared emission using just one or two of the {\em Spitzer} bands, with a strong focus on deriving the total infrared flux from just the higher resolution, shorter wavelength 8~$\mu$m and 24~$\mu$m bands. To do so, we use regions within a subset of nearby face-on spiral galaxies from the Spitzer Nearby Galaxies Survey \citep[SINGS,][]{kennicutt2003a} that are resolved in the {\em Spitzer} 160~$\mu$m band as well as integrated galaxy luminosities for galaxies in the SINGS, the Local Volume Legacy survey (LVL) and the \cite{engelbracht2008a}, hereafter E08, samples. In section \ref{sec:sample}, we describe the samples of galaxies and the data processing. In section \ref{sec:results}, we present the results, and we discuss them in section \ref{sec:discussion}. Finally we summarize our results and conclude in section \ref{sec:conclusion}.

\section{Sample and data}
\label{sec:sample}

We use several samples to derive the TIR emission both from galaxy subregions and galaxies. We use the subset of SINGS galaxies selected by \cite{bendo2008a} to study galaxy subregions. The requirements to perform this study are similar to theirs: the galaxies must have major axes of 5\arcmin\ or greater so that galaxy substructure is resolved; the inclination must be no more than 60$^o$ from face-on or less so that spatial variations can be seen with little overlap due to projection effects; and spatially resolved oxygen abundance data must be available. Galaxies which have compact infrared emission (NGC~1512, NGC~4826 [M64]), where muxbleed effects at 8.0~$\mu$m caused problems with interpreting the data (NGC~1097, NGC~1566 and NGC~4736 [M94]), or where very bright foreground stars caused problems with interpreting the IRAC data (NGC~3561) were excluded. NGC~3938 and NGC~4579 (M58), which were in the \cite{bendo2008a} sample, were not used here. The optical spectrum of NGC~3938 is very noisy, which makes the oxygen abundance measurements unreliable, while the optical spectrum of NGC~4579 is strongly affected by an AGN that causes problems when determining the oxygen abundances (Moustakas et al. 2010, in preparation). The final sample consists of 13 galaxies: NGC~0628 (M74), NGC~0925, NGC~2403, NGC~3031 (M81), NGC~3184, NGC~3351 (M95), NGC~4254 (M99), NGC~4321 (M100), NGC~4725, NGC~5055 (M63), NGC~5194 (M51), NGC~6946 and NGC~7793. 

\begin{figure}[!ht]
\includegraphics[width=\columnwidth]{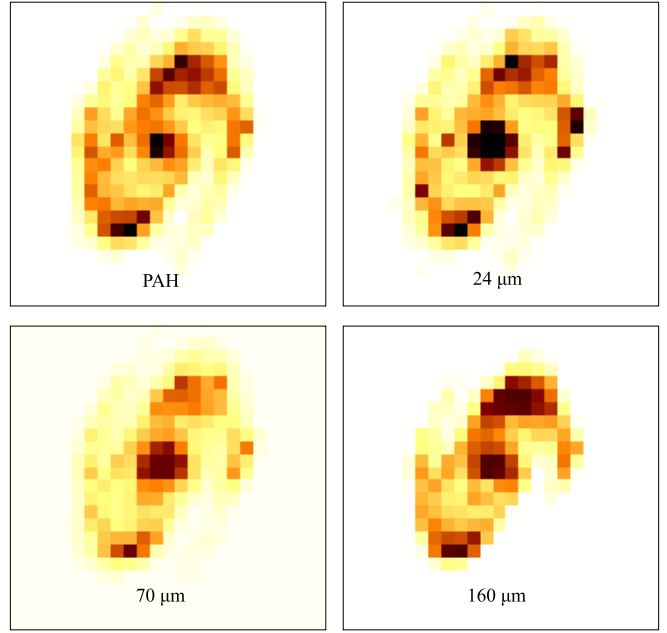}
\caption{Processed PAH 8, 24, 70, and 160~$\mu$m images of NGC~3031. The intensity of the color is proportional to the flux. Each pixel has an angular resolution of 45$\arcsec$. North is up and east is left.\label{fig:M81}}
\end{figure}

Observations and data processing information are provided by \citet{bendo2008a}. The extraction of flux densities for subregions in these galaxies was performed using the method described by \cite{bendo2008a}. First, the data were convolved with kernels developed by \cite{gordon2008a} to match their PSF to that of the 160~$\mu$m data. The entire analysis is performed on the data smoothed to the 160~$\mu$m image resolution. Next the stellar continuum emission was subtracted from the 8.0~$\mu$m and 24~$\mu$m bands using the relations
\begin{equation}
L_\nu\left(PAH~8~\mu m\right)=L_\nu\left(8~\mu m\right)
-0.232\times L_\nu\left(3.6~\mu m\right)
\end{equation}
\begin{equation}
L_\nu\left(24~\mu m\right)=L_\nu\left(24~\mu
 m~obs\right)-0.032\times L_\nu\left(3.6~\mu m\right)
\end{equation}
given by \cite{helou2004a}. The images were then rebinned into 45\arcsec\ pixels, and pixels with low S/N and pixels strongly affected by artifacts or foreground stars were masked out. The processed images of NGC~3031 are shown in Figure~\ref{fig:M81} as an example. The flux densities in these 45\arcsec\ pixels are then used in the analysis. See \cite{bendo2008a} for additional details. The physical resolution ranges from 0.7~kpc to 3.6~kpc with a mean of $2.0\pm1.0$~kpc. The oxygen abundances have been calculated for each 45\arcsec\ region using the abundance gradients from Moustakas et al. (2010, in preparation), assuming that the gradient in each galaxy is azimuthally symmetric. They range from 8.29 to 8.93 with $\left<12+\log O/H\right> = 8.60\pm0.13$ averaging the estimates from the \cite{kobulnicky2004a} (hereafter KK04) and the \cite{pilyugin2005a} (hereafter PT05) estimators. Indeed these estimators are representative of the maximum and minimum oxygen abundance and as such bracket the actual abundance. A similar method was used for the same reasons by \cite{calzetti2007a}. Unless specified otherwise we use this average throughout the paper to estimate the metallicity of galaxy subregions.

We use integrated galaxy data from the SINGS \citep{dale2007a} and LVL \citep{lee2008a,dale2009a} samples as well as the E08 sample as three additional data sets in this analysis. Although only half of the LVL galaxies currently have oxygen abundances available (Marble et al. 2010, submitted), these galaxies along with the SINGS and E08 galaxies allow us to probe a larger range of metallicities than the SINGS galaxies subregions. The SINGS sample contains galaxies with a metallicity from 8.02 to 8.99 with $\left<12+\log O/H\right> = 8.64\pm0.21$ averaging the KK04 and PT05 estimators \citep{calzetti2007a}. Finally, the E08 sample ``includes well-known starbursting or star-forming galaxies from the literature''. It probes a large range of oxygen abundances, from 7.31 to 8.85, using electron temperature measurements and O3N2 \citep{pettini2004a} for the most metal-rich galaxies. The sampling is not even and is mainly concentrated on lower-mid abundances: $\left<12+\log O/H\right> = 8.28\pm0.33$ \citep{engelbracht2008a}. Galaxies which were not detected in one or more bands are excluded from this study. In addition, all galaxies are selected with S/N$>$5. The final sample contains 57 of the 75 SINGS galaxies, 179 of the 258 LVL galaxies and 48 of the 66 E08 galaxies.

\section{Methods to derive the relations}
\label{sec:methods}
As a first step, we estimate the total infrared luminosity from the PAH 8~$\mu$m to 160~$\mu$m Spitzer bands. To do so we use equation 22 from \cite{draine2007a} -- a newer derivation than \cite{dale2002a} -- based on models that have been calibrated using results from {\em Spitzer}:
\begin{equation}
\label{eqn:draine}
\begin{array}{rcl}
L\left(TIR\right)&=&0.95 L\left(PAH~8\mu m\right)+1.15 L\left(24~\mu m\right)\\&&+L\left(70~\mu m\right)+L\left(160~\mu m\right),
\end{array}
\end{equation}
where $L=\nu L_\nu$. The difference with equation 4 of \cite{dale2002a} is minimal, $0.03\pm0.03$ dex (Figure~\ref{fig:metal-draine-dale} and section \ref{sec:discussion}).

To derive the relations to determine the total infrared luminosity from the combination of {\em Spitzer} bands we proceed in two ways. The first way is to perform a linear fit on the colors. For instance we calculate the coefficients $a$ and $b$ of a relation of the form: $\log\left(L(TIR)/L(24~\mu m)\right)=a+b\times\log\left(L(PAH~8~\mu m)/L(24~\mu m)\right)$. For easier use, we provide the relations under the form: $\log L(TIR)=\log L(24~\mu m)+a+b\times\log\left(L(PAH~8~\mu m)/L(24~\mu m)\right)$. We use distance independent quantities -- the ratio of two luminosities -- in order to avoid the well known correlation bias induced by luminosity versus luminosity relations. Indeed luminosity versus luminosity correlations are observed whether or not the data are actually correlated due to the multiplication of the flux density by the square of the distance. The drawback of such a relation is that the luminosities in the two {\em Spitzer} bands are not taken into account independently but are tied through the parameter $b$. To alleviate this limitation, we have performed fits in another distance independent way using the luminosity per unit area $\Sigma$. The relations obtained are of the form: $\log \Sigma(TIR)=a+b\times\log \Sigma(PAH~8~\mu m)+c\times\log \Sigma(24~\mu m)$ for instance. One drawback of this formulation is that it requires the target galaxy to be resolved -- which is often not the case for distant galaxies, especially in mid- and far- infrared bands -- and to know its distance.

In addition to deriving the relations, we also study the influence of the metallicity and the star formation intensity (by way of the total infrared luminosity per unit area). To do so we perform the aforementioned fit dividing the sample into 5 bins each containing the same number of data points in $12+\log O/H$ or $\Sigma(TIR)$. Evaluating the slope change as a function of the abundance allows us to study in detail its influence on the determination of the TIR luminosity. Finally, we also provide relations of the form $\log \Sigma(TIR)=a+b\times\log \Sigma(PAH~8~\mu m)+c\times\log \Sigma(24~\mu m)+d\times\left(12+\log O/H\right)$ to correct for the effect of the metallicity on the estimation of the total infrared luminosity per unit area.

\section{Results}
\label{sec:results}

The relations obtained through the methods described in the previous section are listed in Tables~\ref{tab:relations}, \ref{tab:relations-sigma} and \ref{tab:relations-sigma-oxy}. We present the results hereafter.

\begin{deluxetable*}{cccccc}
\tablecolumns{6} \tablewidth{0pc} \tablecaption{TIR estimations from luminosities\label{tab:relations}}
\tablehead{\colhead{y}&\colhead{$x_1$}&\colhead{$x_2$}&\colhead{a}&\colhead{b}&\colhead{$\sigma$}}
\startdata
$L\left(TIR_{SUB}\right)$&$L\left(24~\mu m\right)$&$L\left(PAH~8\mu m\right)/L\left(24~\mu m\right)$&$1.012\pm0.009$&$0.510\pm0.021$&$0.096$\\
$L\left(TIR_{SINGS}\right)$&$L\left(24~\mu m\right)$&$L\left(PAH~8\mu m\right)/L\left(24~\mu m\right)$&$0.997\pm0.023$&$0.494\pm0.071$&$0.142$\\
$L\left(TIR_{LVL}\right)$&$L\left(24~\mu m\right)$&$L\left(PAH~8\mu m\right)/L\left(24~\mu m\right)$&$1.141\pm0.010$&$0.120\pm0.024$&$0.132$\\
$L\left(TIR_{E08}\right)$&$L\left(24~\mu m\right)$&$L\left(PAH~8\mu m\right)/L\left(24~\mu m\right)$&$0.887\pm0.034$&$0.400\pm0.074$&$0.185$\\
$L\left(TIR_{SUB}\right)$&\nodata&$L\left(PAH~8\mu m\right)$&$4.580\pm0.108$&$0.888\pm0.003$&$0.080$\\
$L\left(TIR_{SINGS}\right)$&\nodata&$L\left(PAH~8\mu m\right)$&$6.696\pm0.532$&$0.836\pm0.015$&$0.113$\\ 
$L\left(TIR_{LVL}\right)$&\nodata&$L\left(PAH~8\mu m\right)$&$8.519\pm0.504$&$0.782\pm0.015$&$0.218$\\
$L\left(TIR_{E08}\right)$&\nodata&$L\left(PAH~8\mu m\right)$&$7.075\pm0.790$&$0.829\pm0.022$&$0.182$\\
$L\left(TIR_{SUB}\right)$&\nodata&$L\left(24~\mu m\right)$&$4.961\pm0.109$&$0.887\pm0.003$&$0.082$\\
$L\left(TIR_{SINGS}\right)$&\nodata&$L\left(24~\mu m\right)$&$3.810\pm0.975$&$0.923\pm0.028$&$0.194$\\
$L\left(TIR_{LVL}\right)$&\nodata&$L\left(24~\mu m\right)$&$1.838\pm0.326$&$0.979\pm0.010$&$0.157$\\
$L\left(TIR_{E08}\right)$&\nodata&$L\left(24~\mu m\right)$&$3.487\pm1.022$&$0.924\pm0.029$&$0.212$\\\hline
$L\left(TIR_{SUB}\right)$&$L\left(70~\mu m\right)$&$L\left(PAH~8~\mu m\right)/L\left(70~\mu m\right)$&$0.662\pm0.003$&$0.423\pm0.010$&$0.059$\\
$L\left(TIR_{SINGS}\right)$&$L\left(70~\mu m\right)$&$L\left(PAH~8~\mu m\right)/L\left(70~\mu m\right)$&$0.583\pm0.017$&$0.300\pm0.032$&$0.063$\\
$L\left(TIR_{LVL}\right)$&$L\left(70~\mu m\right)$&$L\left(PAH~8~\mu m\right)/L\left(70~\mu m\right)$&$0.486\pm0.010$&$0.158\pm0.011$&$0.067$\\
$L\left(TIR_{E08}\right)$&$L\left(70~\mu m\right)$&$L\left(PAH~8~\mu m\right)/L\left(70~\mu m\right)$&$0.478\pm0.018$&$0.198\pm0.023$&$0.055$\\
$L\left(TIR_{SUB}\right)$&$L\left(70~\mu m\right)$&$L\left(24~\mu m\right)/L\left(70~\mu m\right)$&$0.789\pm0.013$&$0.351\pm0.020$&$0.083$\\
$L\left(TIR_{SINGS}\right)$&$L\left(70~\mu m\right)$&$L\left(24~\mu m\right)/L\left(70~\mu m\right)$&$0.497\pm0.050$&$0.078\pm0.075$&$0.101$\\
$L\left(TIR_{LVL}\right)$&$L\left(70~\mu m\right)$&$L\left(24~\mu m\right)/L\left(70~\mu m\right)$&$0.583\pm0.034$&$0.287\pm0.043$&$0.089$\\
$L\left(TIR_{E08}\right)$&$L\left(70~\mu m\right)$&$L\left(24~\mu m\right)/L\left(70~\mu m\right)$&$0.415\pm0.020$&$0.183\pm0.039$&$0.073$\\
$L\left(TIR_{SUB}\right)$&\nodata&$L\left(70~\mu m\right)$&$1.210\pm0.142$&$0.981\pm0.004$&$0.095$\\
$L\left(TIR_{SINGS}\right)$&\nodata&$L\left(70~\mu m\right)$&$-1.057\pm0.580$&$1.042\pm0.016$&$0.101$\\
$L\left(TIR_{LVL}\right)$&\nodata&$L\left(70~\mu m\right)$&$-1.210\pm0.222$&$1.045\pm0.006$&$0.076$\\
$L\left(TIR_{E08}\right)$&\nodata&$L\left(70~\mu m\right)$&$-1.098\pm0.409$&$1.040\pm0.011$&$0.078$\\\hline
$L\left(TIR_{SUB}\right)$&$L\left(160~\mu m\right)$&$L\left(PAH~8~\mu m\right)/L\left(160~\mu m\right)$&$0.454\pm0.005$&$0.283\pm0.010$&$0.050$\\
$L\left(TIR_{SINGS}\right)$&$L\left(160~\mu m\right)$&$L\left(PAH~8~\mu m\right)/L\left(160~\mu m\right)$&$0.599\pm0.037$&$0.370\pm0.069$&$0.125$\\
$L\left(TIR_{LVL}\right)$&$L\left(160~\mu m\right)$&$L\left(PAH~8~\mu m\right)/L\left(160~\mu m\right)$&$0.458\pm0.019$&$0.027\pm0.023$&$0.118$\\
$L\left(TIR_{E08}\right)$&$L\left(160~\mu m\right)$&$L\left(PAH~8~\mu m\right)/L\left(160~\mu m\right)$&$0.831\pm0.038$&$0.310\pm0.083$&$0.182$\\
$L\left(TIR_{SUB}\right)$&$L\left(160~\mu m\right)$&$L\left(24~\mu m\right)/L\left(160~\mu m\right)$&$0.616\pm0.005$&$0.332\pm0.006$&$0.031$\\
$L\left(TIR_{SINGS}\right)$&$L\left(160~\mu m\right)$&$L\left(24~\mu m\right)/L\left(160~\mu m\right)$&$0.714\pm0.014$&$0.436\pm0.019$&$0.047$\\
$L\left(TIR_{LVL}\right)$&$L\left(160~\mu m\right)$&$L\left(24~\mu m\right)/L\left(160~\mu m\right)$&$0.751\pm0.013$&$0.447\pm0.017$&$0.055$\\
$L\left(TIR_{E08}\right)$&$L\left(160~\mu m\right)$&$L\left(24~\mu m\right)/L\left(160~\mu m\right)$&$0.748\pm0.010$&$0.465\pm0.024$&$0.070$\\
$L\left(TIR_{SUB}\right)$&$L\left(160~\mu m\right)$&$L\left(70~\mu m\right)/L\left(160~\mu m\right)$&$0.408\pm0.002$&$0.390\pm0.006$&$0.030$\\
$L\left(TIR_{SINGS}\right)$&$L\left(160~\mu m\right)$&$L\left(70~\mu m\right)/L\left(160~\mu m\right)$&$0.437\pm0.007$&$0.623\pm0.030$&$0.052$\\
$L\left(TIR_{LVL}\right)$&$L\left(160~\mu m\right)$&$L\left(70~\mu m\right)/L\left(160~\mu m\right)$&$0.395\pm0.003$&$0.552\pm0.015$&$0.040$\\
$L\left(TIR_{E08}\right)$&$L\left(160~\mu m\right)$&$L\left(70~\mu m\right)/L\left(160~\mu m\right)$&$0.393\pm0.025$&$0.854\pm0.054$&$0.054$\\
$L\left(TIR_{SUB}\right)$&\nodata&$L\left(160~\mu m\right)$&$-1.278\pm0.093$&$1.047\pm0.003$&$0.057$\\
$L\left(TIR_{SINGS}\right)$&\nodata&$L\left(160~\mu m\right)$&$1.683\pm0.851$&$0.965\pm0.024$&$0.159$\\
$L\left(TIR_{LVL}\right)$&\nodata&$L\left(160~\mu m\right)$&$1.642\pm0.263$&$0.965\pm0.008$&$0.114$\\
$L\left(TIR_{E08}\right)$&\nodata&$L\left(160~\mu m\right)$&$1.747\pm1.004$&$0.971\pm0.028$&$0.215$
\enddata \tablecomments{Coefficient for the fit $\log y=\log x_1+a+b\times \log x_2$. $\sigma$ is the standard deviation of the data points around the best fit line. The subscripts refer to the sample.}
\end{deluxetable*}

\begin{deluxetable*}{ccccccc}
\tablecolumns{7} \tablewidth{0pc} \tablecaption{TIR estimations from luminosities per unit area\label{tab:relations-sigma}}
\tablehead{\colhead{y}&\colhead{$x_1$}&\colhead{$x_2$}&\colhead{a}&\colhead{b}&\colhead{c}&\colhead{$\sigma$}}
\startdata
$\Sigma\left(TIR_{SUB}\right)$&$\Sigma\left(PAH~8\mu m\right)$&$\Sigma\left(24~\mu m\right)$&     $ 5.692\pm 0.124$&$ 0.433\pm 0.014$&$ 0.425\pm 0.014$&$ 0.062$     \\
$\Sigma\left(TIR_{SINGS}\right)$&$\Sigma\left(PAH~8\mu m\right)$&$\Sigma\left(24~\mu m\right)$&   $ 2.085\pm 0.380$&$ 0.525\pm 0.067$&$ 0.442\pm 0.070$&$ 0.133$     \\
$\Sigma\left(TIR_{LVL}\right)$&$\Sigma\left(PAH~8\mu m\right)$&$\Sigma\left(24~\mu m\right)$&     $ 5.408\pm 0.416$&$ 0.174\pm 0.020$&$ 0.693\pm 0.026$&$ 0.105$     \\
$\Sigma\left(TIR_{E08}\right)$&$\Sigma\left(PAH~8\mu m\right)$&$\Sigma\left(24~\mu m\right)$&     $ 6.795\pm 0.780$&$ 0.257\pm 0.053$&$ 0.566\pm 0.050$&$ 0.122$     \\
$\Sigma\left(TIR_{SUB}\right)$&$\Sigma\left(PAH~8\mu m\right)$&\nodata&                           $ 5.982\pm 0.170$&$ 0.845\pm 0.005$&\nodata&$ 0.086$     \\
$\Sigma\left(TIR_{SINGS}\right)$&$\Sigma\left(PAH~8\mu m\right)$&\nodata&                         $ 2.766\pm 0.476$&$ 0.944\pm 0.014$&\nodata&$ 0.175$     \\ 
$\Sigma\left(TIR_{LVL}\right)$&$\Sigma\left(PAH~8\mu m\right)$&\nodata&                           $13.085\pm 0.663$&$ 0.629\pm 0.021$&\nodata&$ 0.233$     \\
$\Sigma\left(TIR_{E08}\right)$&$\Sigma\left(PAH~8\mu m\right)$&\nodata&                           $ 7.771\pm 1.515$&$ 0.799\pm 0.045$&\nodata&$ 0.242$     \\
$\Sigma\left(TIR_{SUB}\right)$&$\Sigma\left(24~\mu m\right)$&\nodata&                             $ 6.418\pm 0.169$&$ 0.842\pm 0.005$&\nodata&$ 0.087$     \\
$\Sigma\left(TIR_{SINGS}\right)$&$\Sigma\left(24~\mu m\right)$&\nodata&                           $ 1.714\pm 0.540$&$ 0.981\pm 0.016$&\nodata&$ 0.192$     \\
$\Sigma\left(TIR_{LVL}\right)$&$\Sigma\left(24~\mu m\right)$&\nodata&                             $ 4.220\pm 0.466$&$ 0.904\pm 0.014$&\nodata&$ 0.127$     \\
$\Sigma\left(TIR_{E08}\right)$&$\Sigma\left(24~\mu m\right)$&\nodata&                             $ 8.083\pm 0.898$&$ 0.783\pm 0.027$&\nodata&$ 0.151$     \\\hline
$\Sigma\left(TIR_{SUB}\right)$&$\Sigma\left(PAH~8~\mu m\right)$&$\Sigma\left(70~\mu m\right)$&    $ 3.994\pm 0.070$&$ 0.425\pm 0.006$&$ 0.476\pm 0.006$&$ 0.033$     \\
$\Sigma\left(TIR_{SINGS}\right)$&$\Sigma\left(PAH~8~\mu m\right)$&$\Sigma\left(70~\mu m\right)$&  $ 0.627\pm 0.207$&$ 0.302\pm 0.034$&$ 0.697\pm 0.037$&$ 0.063$     \\
$\Sigma\left(TIR_{LVL}\right)$&$\Sigma\left(PAH~8~\mu m\right)$&$\Sigma\left(70~\mu m\right)$&    $ 1.978\pm 0.295$&$ 0.178\pm 0.011$&$ 0.777\pm 0.016$&$ 0.063$     \\
$\Sigma\left(TIR_{E08}\right)$&$\Sigma\left(PAH~8~\mu m\right)$&$\Sigma\left(70~\mu m\right)$&    $ 0.085\pm 0.433$&$ 0.194\pm 0.023$&$ 0.817\pm 0.028$&$ 0.054$     \\
$\Sigma\left(TIR_{SUB}\right)$&$\Sigma\left(24~\mu m\right)$&$\Sigma\left(70~\mu m\right)$&       $ 4.593\pm 0.137$&$ 0.405\pm 0.015$&$ 0.483\pm 0.016$&$ 0.063$     \\
$\Sigma\left(TIR_{SINGS}\right)$&$\Sigma\left(24~\mu m\right)$&$\Sigma\left(70~\mu m\right)$&     $-0.018\pm 0.319$&$ 0.069\pm 0.076$&$ 0.946\pm 0.078$&$ 0.100$     \\
$\Sigma\left(TIR_{LVL}\right)$&$\Sigma\left(24~\mu m\right)$&$\Sigma\left(70~\mu m\right)$&       $ 1.078\pm 0.400$&$ 0.310\pm 0.045$&$ 0.675\pm 0.049$&$ 0.089$     \\
$\Sigma\left(TIR_{E08}\right)$&$\Sigma\left(24~\mu m\right)$&$\Sigma\left(70~\mu m\right)$&       $ 1.260\pm 0.698$&$ 0.216\pm 0.047$&$ 0.760\pm 0.061$&$ 0.072$     \\
$\Sigma\left(TIR_{SUB}\right)$&$\Sigma\left(70~\mu m\right)$&\nodata&                             $ 3.847\pm 0.176$&$ 0.902\pm 0.005$&\nodata&$ 0.082$     \\
$\Sigma\left(TIR_{SINGS}\right)$&$\Sigma\left(70~\mu m\right)$&\nodata&                           $-0.118\pm 0.299$&$ 1.017\pm 0.009$&\nodata&$ 0.101$     \\
$\Sigma\left(TIR_{LVL}\right)$&$\Sigma\left(70~\mu m\right)$&\nodata&                             $ 0.042\pm 0.417$&$ 1.010\pm 0.013$&\nodata&$ 0.100$     \\
$\Sigma\left(TIR_{E08}\right)$&$\Sigma\left(70~\mu m\right)$&\nodata&                             $-0.617\pm 0.673$&$ 1.028\pm 0.020$&\nodata&$ 0.087$     \\\hline
$\Sigma\left(TIR_{SUB}\right)$&$\Sigma\left(PAH~8~\mu m\right)$&$\Sigma\left(160~\mu m\right)$&   $-1.794\pm 0.180$&$ 0.156\pm 0.014$&$ 0.908\pm 0.018$&$ 0.046$     \\
$\Sigma\left(TIR_{SINGS}\right)$&$\Sigma\left(PAH~8~\mu m\right)$&$\Sigma\left(160~\mu m\right)$& $ 1.474\pm 0.365$&$ 0.408\pm 0.068$&$ 0.567\pm 0.071$&$ 0.119$     \\
$\Sigma\left(TIR_{LVL}\right)$&$\Sigma\left(PAH~8~\mu m\right)$&$\Sigma\left(160~\mu m\right)$&   $ 0.957\pm 0.632$&$ 0.040\pm 0.028$&$ 0.946\pm 0.042$&$ 0.118$     \\
$\Sigma\left(TIR_{E08}\right)$&$\Sigma\left(PAH~8~\mu m\right)$&$\Sigma\left(160~\mu m\right)$&   $ 0.396\pm 1.709$&$ 0.300\pm 0.092$&$ 0.713\pm 0.122$&$ 0.182$     \\
$\Sigma\left(TIR_{SUB}\right)$&$\Sigma\left(24~\mu m\right)$&$\Sigma\left(160~\mu m\right)$&      $-0.641\pm 0.094$&$ 0.283\pm 0.006$&$ 0.753\pm 0.008$&$ 0.028$     \\
$\Sigma\left(TIR_{SINGS}\right)$&$\Sigma\left(24~\mu m\right)$&$\Sigma\left(160~\mu m\right)$&    $ 0.823\pm 0.134$&$ 0.435\pm 0.019$&$ 0.562\pm 0.019$&$ 0.046$     \\
$\Sigma\left(TIR_{LVL}\right)$&$\Sigma\left(24~\mu m\right)$&$\Sigma\left(160~\mu m\right)$&      $ 1.177\pm 0.228$&$ 0.452\pm 0.017$&$ 0.535\pm 0.019$&$ 0.055$     \\
$\Sigma\left(TIR_{E08}\right)$&$\Sigma\left(24~\mu m\right)$&$\Sigma\left(160~\mu m\right)$&      $ 1.955\pm 0.602$&$ 0.482\pm 0.025$&$ 0.482\pm 0.035$&$ 0.067$     \\
$\Sigma\left(TIR_{SUB}\right)$&$\Sigma\left(70~\mu m\right)$&$\Sigma\left(160~\mu m\right)$&      $-1.357\pm 0.070$&$ 0.325\pm 0.006$&$ 0.727\pm 0.007$&$ 0.024$     \\
$\Sigma\left(TIR_{SINGS}\right)$&$\Sigma\left(70~\mu m\right)$&$\Sigma\left(160~\mu m\right)$&    $ 0.051\pm 0.153$&$ 0.634\pm 0.030$&$ 0.377\pm 0.030$&$ 0.051$     \\
$\Sigma\left(TIR_{LVL}\right)$&$\Sigma\left(70~\mu m\right)$&$\Sigma\left(160~\mu m\right)$&      $-0.798\pm 0.145$&$ 0.571\pm 0.013$&$ 0.465\pm 0.013$&$ 0.034$     \\
$\Sigma\left(TIR_{E08}\right)$&$\Sigma\left(70~\mu m\right)$&$\Sigma\left(160~\mu m\right)$&      $-1.570\pm 0.635$&$ 0.843\pm 0.050$&$ 0.215\pm 0.055$&$ 0.075$     \\
$\Sigma\left(TIR_{SUB}\right)$&$\Sigma\left(160~\mu m\right)$&\nodata&                            $-3.288\pm 0.127$&$ 1.107\pm 0.004$&\nodata&$ 0.049$     \\
$\Sigma\left(TIR_{SINGS}\right)$&$\Sigma\left(160~\mu m\right)$&\nodata&                          $ 0.791\pm 0.438$&$ 0.989\pm 0.013$&\nodata&$ 0.152$     \\
$\Sigma\left(TIR_{LVL}\right)$&$\Sigma\left(160~\mu m\right)$&\nodata&                            $ 0.457\pm 0.496$&$ 0.999\pm 0.015$&\nodata&$ 0.120$     \\
$\Sigma\left(TIR_{E08}\right)$&$\Sigma\left(160~\mu m\right)$&\nodata&                            $-2.001\pm 1.697$&$ 1.081\pm 0.050$&\nodata&$ 0.203$
\enddata \tablecomments{Coefficient for the fit $\log y=a+b\times \log x_1+c\times \log x_2$. $\sigma$ is the standard deviation of the data points around the best fit line. The subscripts refer to the sample.}
\end{deluxetable*}

\begin{deluxetable*}{cccccccc}
\tablecolumns{8} \tablewidth{0pc} \tablecaption{TIR estimations from luminosities per unit area and the oxygen abundance\label{tab:relations-sigma-oxy}}
\tablehead{\colhead{y}&\colhead{$x_1$}&\colhead{$x_2$}&\colhead{a}&\colhead{b}&\colhead{c}&\colhead{d}&\colhead{$\sigma$}}
\startdata
$\Sigma\left(TIR_{SUB}\right)$&$\Sigma\left(PAH~8\mu m\right)$&$\Sigma\left(24~\mu m\right)$&    $ 5.667\pm 0.138$&$ 0.431\pm 0.015$&$ 0.426\pm 0.014$&$ 0.009\pm 0.022$&$ 0.062$\\
$\Sigma\left(TIR_{SINGS}\right)$&$\Sigma\left(PAH~8\mu m\right)$&$\Sigma\left(24~\mu m\right)$&  $ 3.996\pm 0.909$&$ 0.637\pm 0.081$&$ 0.339\pm 0.081$&$-0.256\pm 0.112$&$ 0.127$\\
$\Sigma\left(TIR_{E08}\right)$&$\Sigma\left(PAH~8\mu m\right)$&$\Sigma\left(24~\mu m\right)$&    $ 7.214\pm 0.790$&$ 0.336\pm 0.066$&$ 0.508\pm 0.057$&$-0.134\pm 0.071$&$ 0.118$\\
$\Sigma\left(TIR_{SUB}\right)$&$\Sigma\left(PAH~8\mu m\right)$&\nodata&                          $ 6.122\pm 0.189$&$ 0.854\pm 0.007$&\nodata&$-0.051\pm 0.030$&$ 0.086$\\
$\Sigma\left(TIR_{SINGS}\right)$&$\Sigma\left(PAH~8\mu m\right)$&\nodata&                        $ 6.283\pm 0.828$&$ 0.972\pm 0.013$&\nodata&$-0.514\pm 0.106$&$ 0.146$\\ 
$\Sigma\left(TIR_{E08}\right)$&$\Sigma\left(PAH~8\mu m\right)$&\nodata&                          $ 8.893\pm 1.271$&$ 0.882\pm 0.041$&\nodata&$-0.471\pm 0.099$&$ 0.197$\\
$\Sigma\left(TIR_{SUB}\right)$&$\Sigma\left(24~\mu m\right)$&\nodata&                            $ 5.711\pm 0.186$&$ 0.805\pm 0.007$&\nodata&$ 0.222\pm 0.028$&$ 0.084$\\
$\Sigma\left(TIR_{SINGS}\right)$&$\Sigma\left(24~\mu m\right)$&\nodata&                          $-0.147\pm 1.050$&$ 0.970\pm 0.017$&\nodata&$ 0.260\pm 0.127$&$ 0.185$\\
$\Sigma\left(TIR_{E08}\right)$&$\Sigma\left(24~\mu m\right)$&\nodata&                            $ 7.522\pm 0.982$&$ 0.777\pm 0.027$&\nodata&$ 0.092\pm 0.068$&$ 0.148$\\\hline
$\Sigma\left(TIR_{SUB}\right)$&$\Sigma\left(PAH~8~\mu m\right)$&$\Sigma\left(70~\mu m\right)$&   $ 3.606\pm 0.075$&$ 0.393\pm 0.006$&$ 0.488\pm 0.006$&$ 0.122\pm 0.011$&$ 0.031$\\
$\Sigma\left(TIR_{SINGS}\right)$&$\Sigma\left(PAH~8~\mu m\right)$&$\Sigma\left(70~\mu m\right)$& $ 0.045\pm 0.541$&$ 0.266\pm 0.046$&$ 0.732\pm 0.048$&$ 0.069\pm 0.060$&$ 0.063$\\
$\Sigma\left(TIR_{E08}\right)$&$\Sigma\left(PAH~8~\mu m\right)$&$\Sigma\left(70~\mu m\right)$&   $ 0.690\pm 0.461$&$ 0.241\pm 0.028$&$ 0.773\pm 0.031$&$-0.082\pm 0.030$&$ 0.050$\\
$\Sigma\left(TIR_{SUB}\right)$&$\Sigma\left(24~\mu m\right)$&$\Sigma\left(70~\mu m\right)$&      $ 3.497\pm 0.138$&$ 0.326\pm 0.014$&$ 0.513\pm 0.014$&$ 0.308\pm 0.018$&$ 0.056$\\
$\Sigma\left(TIR_{SINGS}\right)$&$\Sigma\left(24~\mu m\right)$&$\Sigma\left(70~\mu m\right)$&    $-1.941\pm 0.505$&$ 0.054\pm 0.065$&$ 0.949\pm 0.067$&$ 0.268\pm 0.059$&$ 0.086$\\
$\Sigma\left(TIR_{E08}\right)$&$\Sigma\left(24~\mu m\right)$&$\Sigma\left(70~\mu m\right)$&      $ 0.834\pm 0.684$&$ 0.215\pm 0.045$&$ 0.755\pm 0.058$&$ 0.078\pm 0.031$&$ 0.067$\\
$\Sigma\left(TIR_{SUB}\right)$&$\Sigma\left(70~\mu m\right)$&\nodata&                            $ 2.450\pm 0.161$&$ 0.827\pm 0.006$&\nodata&$ 0.453\pm 0.022$&$ 0.069$\\
$\Sigma\left(TIR_{SINGS}\right)$&$\Sigma\left(70~\mu m\right)$&\nodata&                          $-2.037\pm 0.491$&$ 1.004\pm 0.008$&\nodata&$ 0.271\pm 0.059$&$ 0.086$\\
$\Sigma\left(TIR_{E08}\right)$&$\Sigma\left(70~\mu m\right)$&\nodata&                            $-1.042\pm 0.682$&$ 1.021\pm 0.019$&\nodata&$ 0.079\pm 0.038$&$ 0.083$\\\hline
$\Sigma\left(TIR_{SUB}\right)$&$\Sigma\left(PAH~8~\mu m\right)$&$\Sigma\left(160~\mu m\right)$&  $-1.566\pm 0.170$&$ 0.164\pm 0.013$&$ 0.939\pm 0.017$&$-0.178\pm 0.015$&$ 0.043$\\
$\Sigma\left(TIR_{SINGS}\right)$&$\Sigma\left(PAH~8~\mu m\right)$&$\Sigma\left(160~\mu m\right)$&$ 4.522\pm 0.537$&$ 0.476\pm 0.052$&$ 0.519\pm 0.054$&$-0.430\pm 0.065$&$ 0.088$\\
$\Sigma\left(TIR_{E08}\right)$&$\Sigma\left(PAH~8~\mu m\right)$&$\Sigma\left(160~\mu m\right)$&  $ 2.514\pm 1.504$&$ 0.449\pm 0.084$&$ 0.590\pm 0.105$&$-0.357\pm 0.079$&$ 0.151$\\
$\Sigma\left(TIR_{SUB}\right)$&$\Sigma\left(24~\mu m\right)$&$\Sigma\left(160~\mu m\right)$&     $-0.626\pm 0.088$&$ 0.274\pm 0.006$&$ 0.790\pm 0.008$&$-0.115\pm 0.009$&$ 0.026$\\
$\Sigma\left(TIR_{SINGS}\right)$&$\Sigma\left(24~\mu m\right)$&$\Sigma\left(160~\mu m\right)$&   $ 1.463\pm 0.250$&$ 0.419\pm 0.018$&$ 0.582\pm 0.019$&$-0.094\pm 0.032$&$ 0.043$\\
$\Sigma\left(TIR_{E08}\right)$&$\Sigma\left(24~\mu m\right)$&$\Sigma\left(160~\mu m\right)$&     $ 2.020\pm 0.590$&$ 0.472\pm 0.025$&$ 0.504\pm 0.037$&$-0.055\pm 0.032$&$ 0.065$\\
$\Sigma\left(TIR_{SUB}\right)$&$\Sigma\left(70~\mu m\right)$&$\Sigma\left(160~\mu m\right)$&     $-1.350\pm 0.070$&$ 0.328\pm 0.006$&$ 0.721\pm 0.008$&$ 0.011\pm 0.009$&$ 0.024$\\
$\Sigma\left(TIR_{SINGS}\right)$&$\Sigma\left(70~\mu m\right)$&$\Sigma\left(160~\mu m\right)$&   $-0.385\pm 0.329$&$ 0.658\pm 0.034$&$ 0.351\pm 0.034$&$ 0.060\pm 0.040$&$ 0.050$\\
$\Sigma\left(TIR_{E08}\right)$&$\Sigma\left(70~\mu m\right)$&$\Sigma\left(160~\mu m\right)$&     $-1.632\pm 0.645$&$ 0.856\pm 0.054$&$ 0.198\pm 0.061$&$ 0.026\pm 0.038$&$ 0.075$\\
$\Sigma\left(TIR_{SUB}\right)$&$\Sigma\left(160~\mu m\right)$&\nodata&                           $-3.199\pm 0.117$&$ 1.151\pm 0.005$&\nodata&$-0.180\pm 0.013$&$ 0.045$\\
$\Sigma\left(TIR_{SINGS}\right)$&$\Sigma\left(160~\mu m\right)$&\nodata&                         $ 2.848\pm 0.792$&$ 1.005\pm 0.013$&\nodata&$-0.301\pm 0.099$&$ 0.141$\\
$\Sigma\left(TIR_{E08}\right)$&$\Sigma\left(160~\mu m\right)$&\nodata&                           $-1.507\pm 1.659$&$ 1.113\pm 0.051$&\nodata&$-0.189\pm 0.092$&$ 0.194$
\enddata \tablecomments{Coefficient for the fit $\log y=a+b\times \log x_1+c\times \log x_2+d\times\left(12+\log O/H\right)$. $\sigma$ is the standard deviation of the data points around the best fit line. The subscripts refer to the sample.}
\end{deluxetable*}                                                                               

\subsection{TIR estimates from the combination of the 8 and 24~$\mu$m bands}
\label{ssec:8-24}
As PAH and hot dust are significant contributors to the total infrared luminosity in metal-rich galaxies \citep{draine2007a}, the 8~$\mu$m and 24~$\mu$m observations permit an accurate determination of the TIR luminosities. Furthermore, as their spatial resolution is significantly better than the one of the 70~$\mu$m and the 160~$\mu$m bands, resolving local galaxies is possible.


\subsubsection{Metallicity effects\label{sssec:metallicity}}

It is well known that the presence of dust carriers influencing the 8~$\mu$m band emission is directly affected by the metallicity \citep{engelbracht2005a,rosenberg2006a,wu2006a,madden2006a,jackson2006a,draine2007a}. The large number of galaxy subregions in the sample studied here allows us to take into account accurately this parameter in the determination of the total infrared emission from the PAH 8~$\mu$m and 24~$\mu$m bands. In addition, the integrated SINGS and the E08 galaxies have oxygen abundances that are averaged over all the galaxy.

\begin{figure}[!ht]
\includegraphics[width=\columnwidth]{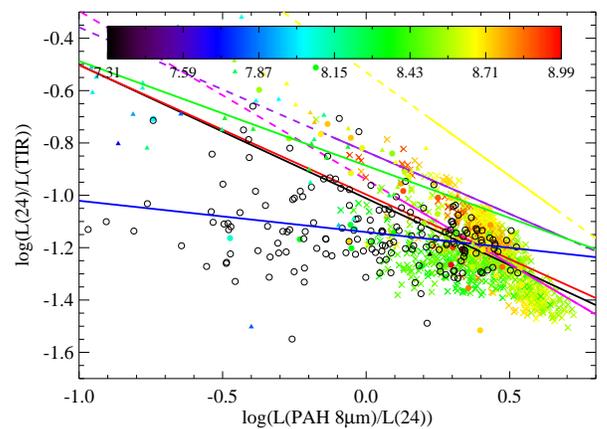}
\caption{Ratio of the 24~$\mu$m to the total infrared emission -- as calculated using equation \ref{eqn:draine} -- versus the ratio of the PAH 8~$\mu$m to 24~$\mu$m emission. The ``x'' symbol represents the data for the subregions in the sample of face-on galaxies. The color of the ``x'' symbols represent the oxygen abundances as given by Moustakas et al. (2010, in preparation). The filled circles represent integrated galaxy data for the SINGS sample, the empty circles represent integrated galaxy data for the LVL sample, and the filled triangles represent integrated galaxy data for the E08 sample. The solid lines represent various relations: the yellow one for the relation found in NGC~5194 by \cite{calzetti2005a}, the magenta one for the relation found in NGC~3031 by \cite{perez2006a}, the purple one for the relation found in M~33 by \cite{thilker2007a}, the blue one for LVL galaxies, the red one for SINGS galaxies, the green one for E08 galaxies, and the black one for galaxy subregions. For the relations published in the literature, the $L(PAH~8~\mu m)/L(24~\mu m)$ range over which they were derived are displayed with a solid line, and a dashed line out of these bounds (P. P\'erez-Gonz\'alez and D. Thilker, private communications).\label{fig:PAH-24-TIR}}
\end{figure}

The relations between the $L(24~\mu m)/L(TIR)$ and $L(PAH~8\mu m)/L(24~\mu m)$ colors for all the samples are presented in Figure~\ref{fig:PAH-24-TIR}. In order to clearly show the difference in the behavior of integrated galaxies and galaxy subregions, we plot them together. For comparison we also show the relations for NGC~5194, NGC~3031, and M33 determined by \cite{calzetti2005a}, \cite{perez2006a}, and \cite{thilker2007a} respectively. The NGC~5194 relation by \cite{calzetti2005a} is offset from the other relations and also from the observations as it has a much higher $L(24~\mu m)/L(TIR)$ ratio. The reason is that the NGC~5194 flux densities were measured after removing a local background, whereas for NGC~3031 and other data, flux densities were measured after removing only a global background. The difference therefore could be an indication of the relative contribution of the diffuse large scale infrared emission compared to point-like sources. The parameters of the best fitting relations are given in Table~\ref{tab:relations}, separately for each sample and the galaxy subregions. In addition, we provide relations between the total infrared luminosity per unit area as a combination of the luminosity per unit area in the PAH 8~$\mu$m and 24~$\mu$m bands in Table~\ref{tab:relations-sigma} and we also take into account the oxygen abundance in Table~\ref{tab:relations-sigma-oxy}.

First of all in Figure~\ref{fig:PAH-24-TIR} we notice that despite the scatter, different relations can be seen for subregions with different abundances. Indeed, we see that the slope is clearly shallower for the lower oxygen abundance subregions which describe a lower branch in the diagram. Interestingly, the low metallicity trend is followed by a part of the SINGS and most of the LVL samples. Even though the LVL sample spans a significant range of metallicities, it is chiefly constituted of dwarf galaxies. As a consequence, LVL galaxies are statistically more metal-poor than SINGS ones and therefore exhibit for a significant number of them little or no PAH emission. So in this particular case the data follow a trend similar to the one for the lowest oxygen abundance galaxy subregions (note that $12+\log O/H>8.29$ for the subregions). For most of the LVL galaxies, the $L(PAH~8\mu m)/L(24~\mu m)$ ratio is primarily dependent on the PAH 8~$\mu$m luminosity which itself is dependent on the metallicity. That is, the $L(24~\mu m)/L(TIR)$ ratio is hardly affected by a change of the $L(PAH~8\mu m)/L(24~\mu m)$ color. Quantitatively, we see in Table~\ref{tab:relations} that the slope is indeed much shallower for SINGS and especially LVL galaxies, which show little dependence on $L(PAH~8\mu m)/L(24~\mu m)$ as could be expected. The subregions sample has the highest mean slope due to being constituted of spiral galaxies only. The behavior of the E08 sample shows that the metallicity is not the only parameter driving the correlation between $L(24~\mu m)/L(TIR)$ and $L(PAH~8\mu m)/L(24~\mu m)$. Most of the sample seems to follow the trend set by higher metallicity subregions describing the upper branch in the diagram. As this sample has been specifically constituted with star-forming and starburst galaxies, the chief parameter that drives the variation could be the intensity of star formation. We will probe its effect in section \ref{sssec:SFR-intensity}.

In Table~\ref{tab:relations-sigma-oxy} we estimate the total infrared luminosity per unit area as a function of the luminosity per unit area in the PAH 8~$\mu$m and 24~$\mu$m bands as well as the oxygen abundance. We see that combining the PAH 8~$\mu$m and 24~$\mu$m bands, the contribution of the metallicity is smaller than when estimating from the PAH 8~$\mu$m band only. The reason is that the combination of these two bands is an indirect measure of the metallicity which permits to obtain a more accurate measurement than when no information on the metallicity is available. Indeed, taking into account the oxygen abundance, averaging over the subregions, SINGS and E08 samples, the scatter is reduced by 0.003 dex when combining the PAH 8~$\mu$m and 24~$\mu$m bands but 0.025 dex when estimating the total infrared emission from the one in the PAH 8~$\mu$m band only.

\begin{figure*}[!ht]                                                                             
\includegraphics[width=\columnwidth]{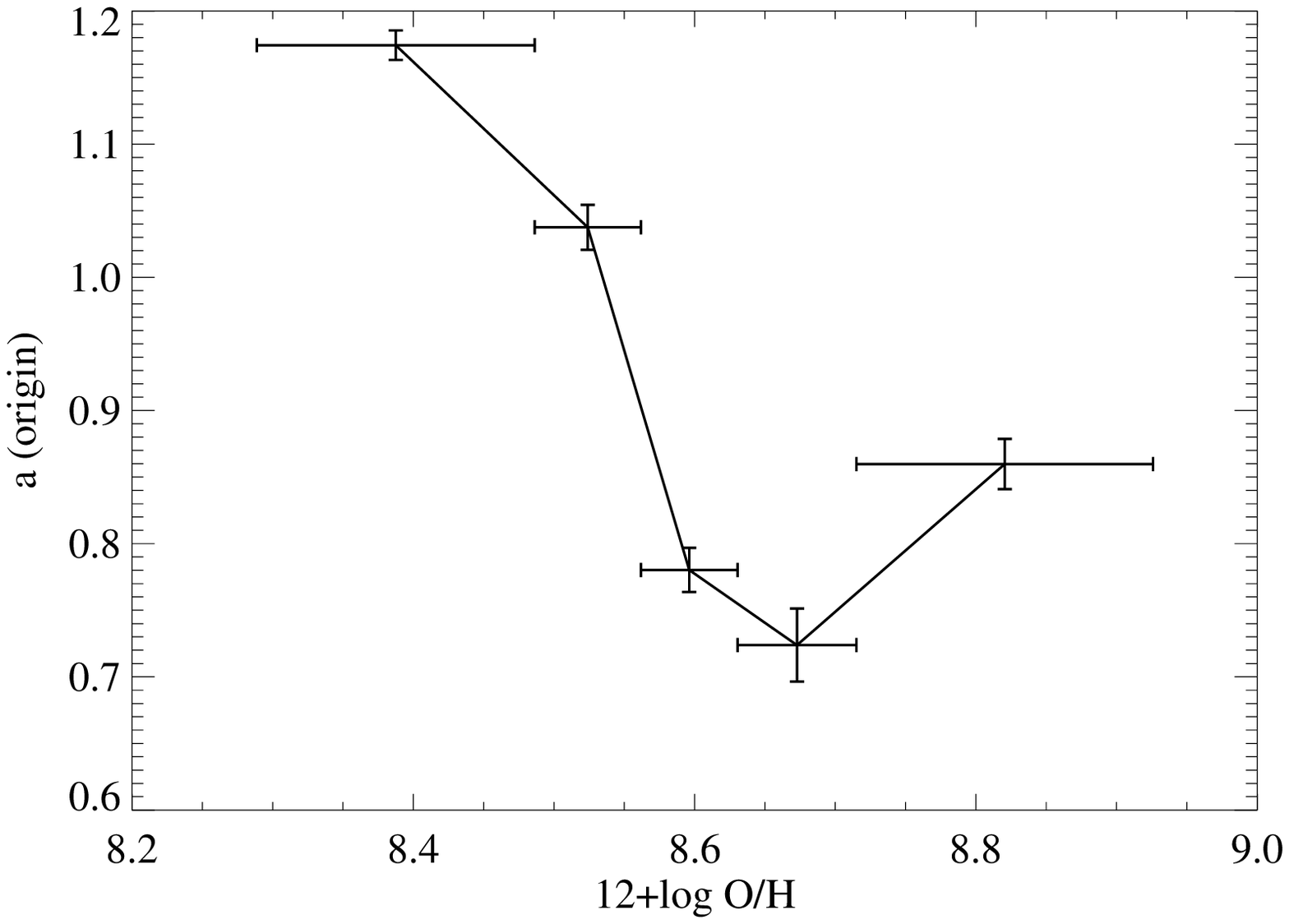}                                          
\includegraphics[width=\columnwidth]{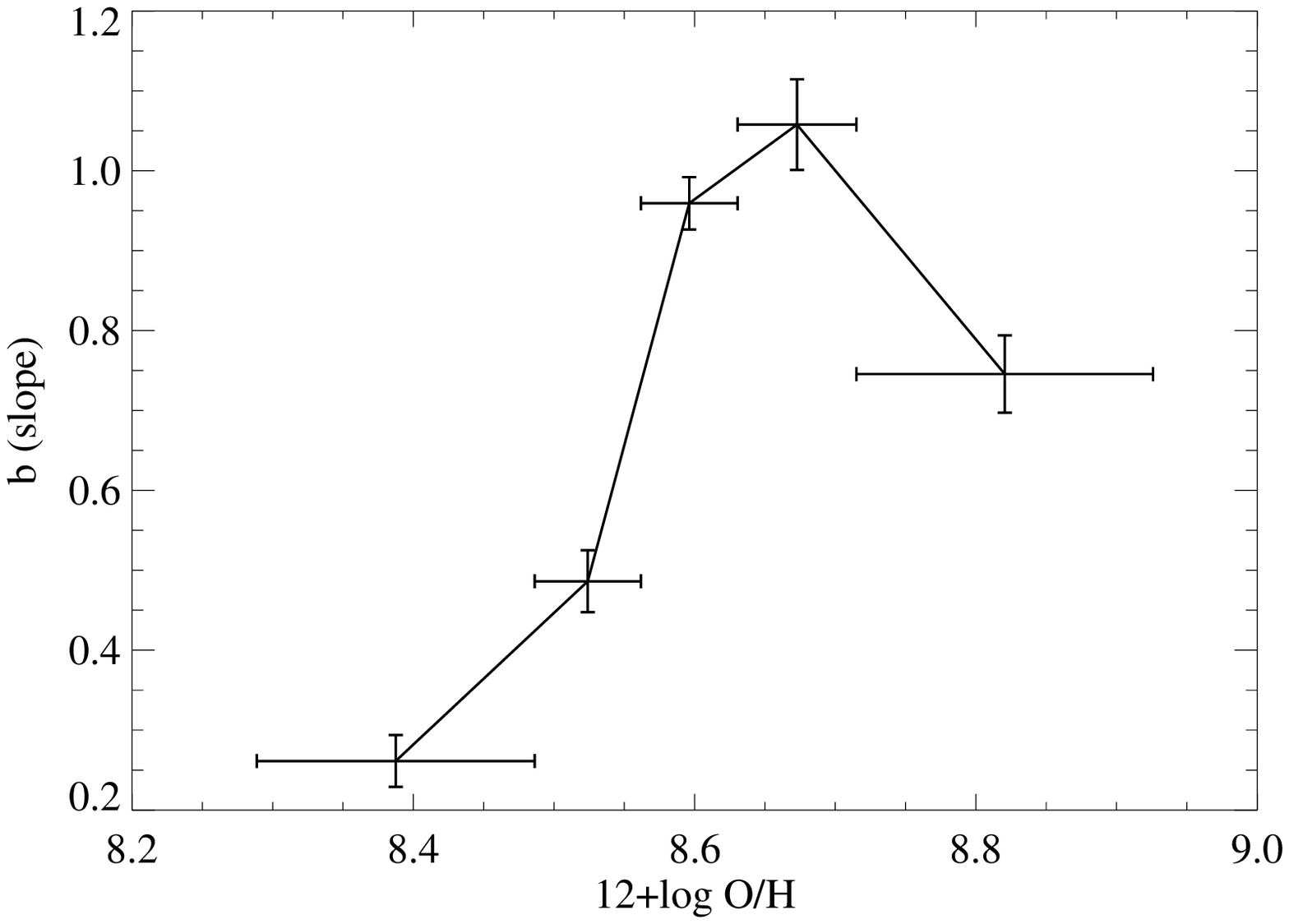}                                           
\caption{Top: y-intercept $a$ (left) and slope $b$ (right) of the linear best fit: $\log L(TIR) = \log L(24~\mu m)+a+b\times\log \left(L(8~\mu m)/L(24~\mu m)\right)$ versus the oxygen abundance. The data points are calculated dividing the sample in 5 bins each containing the same number of data points. Here only subregions are taken into account.\label{fig:fit-subregions}}
\end{figure*}

In order to study how the fit parameters evolve as a function of the oxygen abundance, in Figure~\ref{fig:fit-subregions}, we plot the best fit parameters for different bins of oxygen abundances following the method described earlier in section \ref{sec:methods}.

Three different trends can be seen with increasing oxygen abundance: 1. at low oxygen abundance, the slope is shallow, 2. at a higher oxygen abundance, the slope is steeper, this shows the greater dependence on the PAH emission, 3. finally at the highest oxygen abundance, the slope decreases slightly. Considering that $L\left(24~\mu m\right)/L\left(TIR\right)$ is a function of the intensity of the illuminating radiation field \citep{dale2001a,draine2007a}, the observed trend reflects the results obtained by other authors \citep[e. g.][]{calzetti2005a,bendo2006a,bendo2008a} that the relative strength of PAH emission decreases as the strength of the illuminating radiation field increases. This could be because of PAH destruction or the inhibition of PAH emission at 8~$\mu$m in regions with intense or hard radiation fields. The ratio $L(PAH~8\mu m)/L(24~\mu m)$ may drop off more gradually in high metallicity galaxies because of increased dust extinction. The photons that most strongly affect PAHs in those systems can only travel a short distance in the ISM, so the drop off in $L(PAH~8\mu m)/L(24~\mu m)$ with increasing radiation field intensity is more gradual. In low metallicity systems, however, the photons that affect PAH emission can propagate much further through the ISM, so $L(PAH~8\mu m)/L(24~\mu m)$ decreases much more quickly as the intensity of the radiation field (and $L(24~\mu m)/L(TIR)$) increases.
                                                                                                 
\subsubsection{Star formation intensity effect\label{sssec:SFR-intensity}}                       
                                                                                                 
As mentioned in section \ref{sssec:metallicity}, the metallicity cannot explain that some low-metallicity galaxies follow the trend set by rather high-metallicity subregions. This hints that another parameter is playing a role in driving the correlation.

\begin{figure}[!ht]                                                                              
\includegraphics[width=\columnwidth]{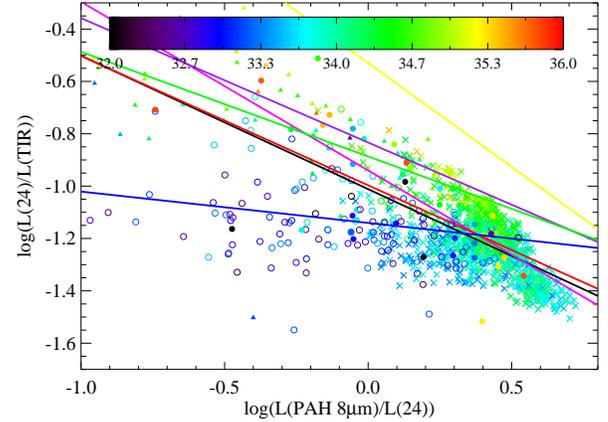}            
\caption{The axis and the shapes of the symbols are the same as in Figure~\ref{fig:PAH-24-TIR}. The symbols are color-coded as a function of $\log \Sigma\left(TIR\right)$ (W\ kpc$^{-2}$).\label{fig:PAH-24-TIR-SFR}}
\end{figure}

\begin{figure*}[!ht]
\includegraphics[width=\columnwidth]{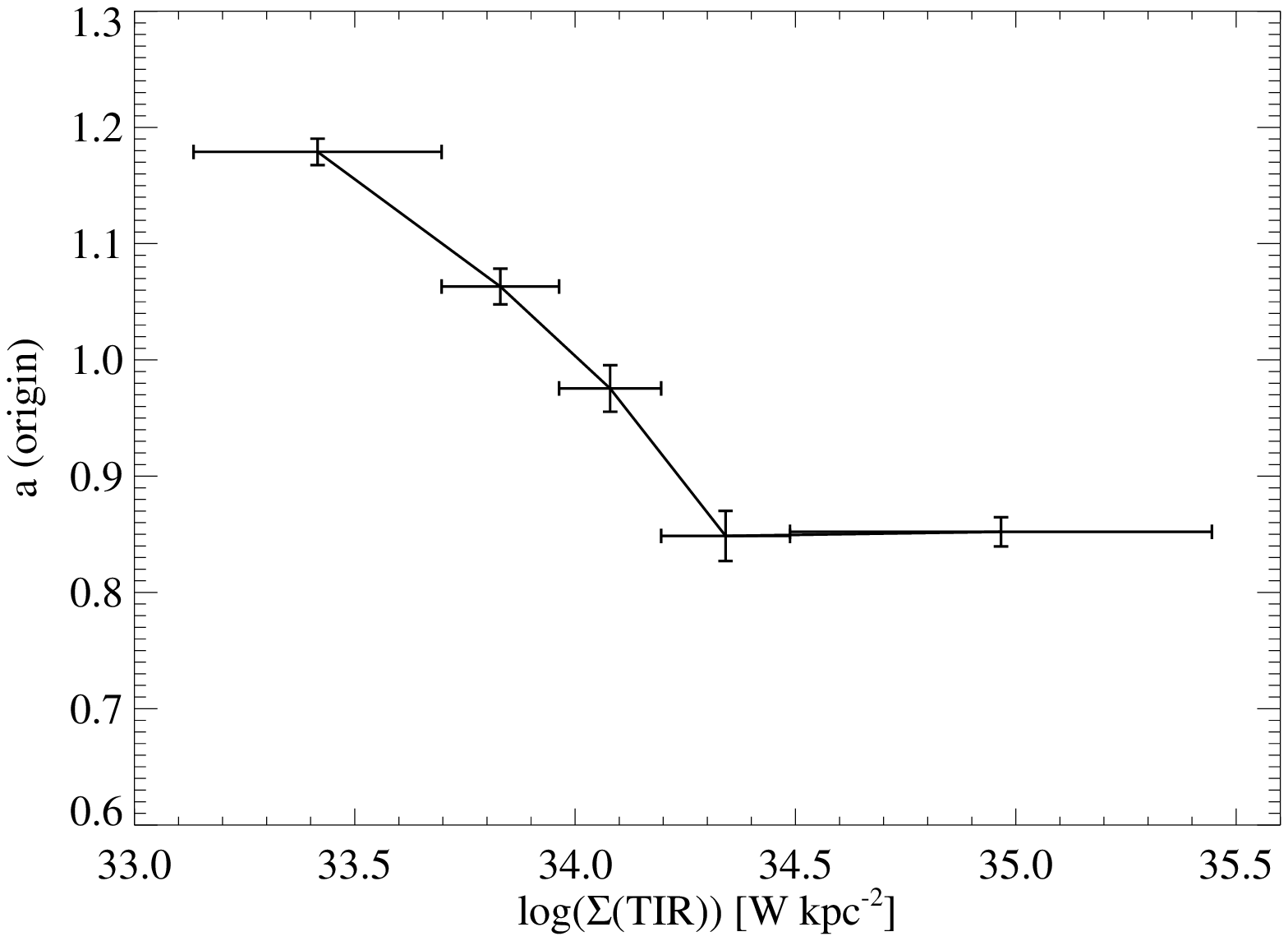}
\includegraphics[width=\columnwidth]{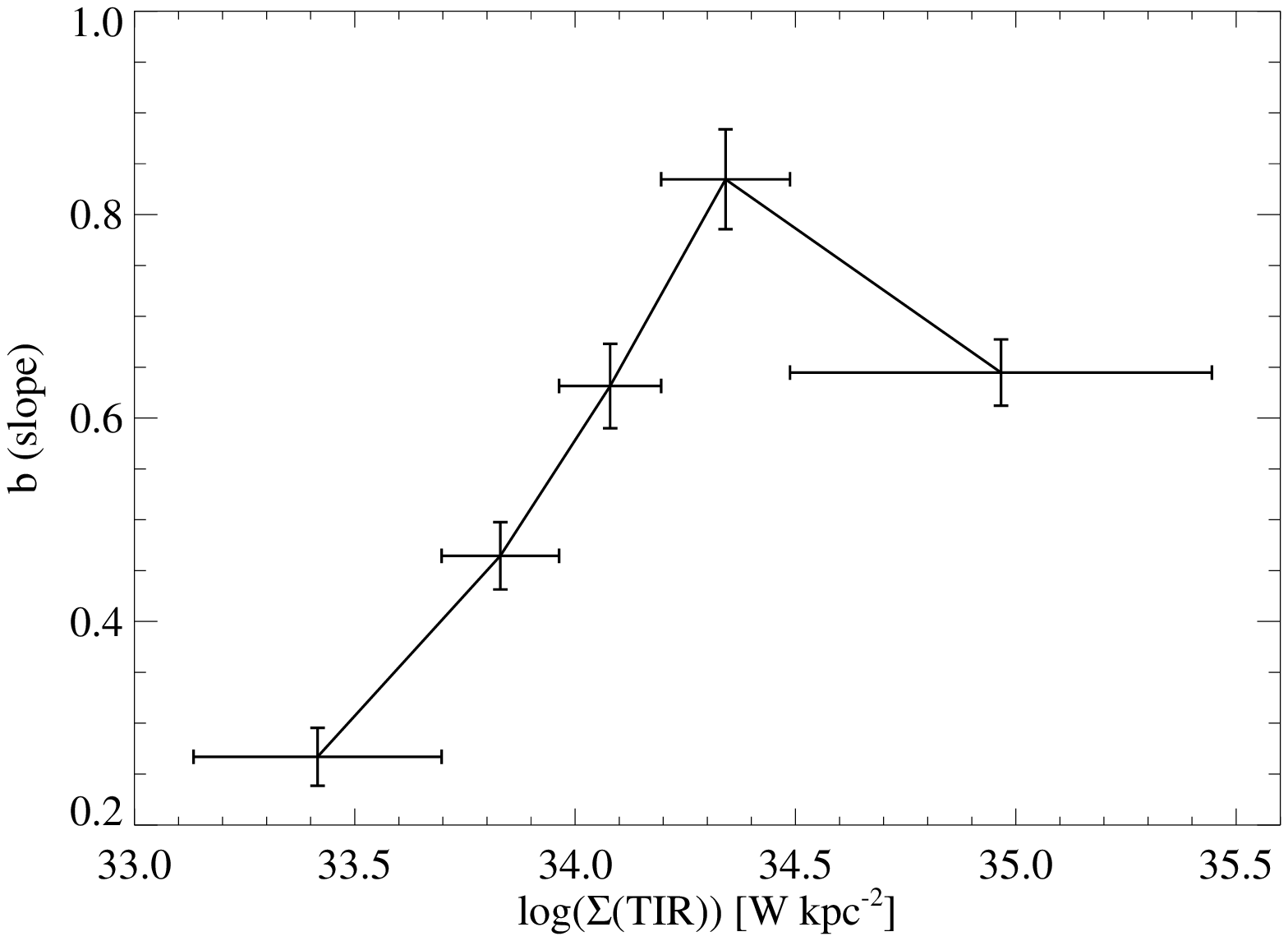}
\caption{y-intercept $a$ (left) and slope $b$ (right) of the linear best fit: $\log L(TIR) = \log L(24~\mu m)+a+b\log\left( L(8~\mu m)/L(24~\mu m)\right)$ versus $\Sigma\left(TIR\right)$.\label{fig:fit-subregions-SFR}}
\end{figure*}

In Figure~\ref{fig:PAH-24-TIR-SFR} we plot the $\log \left(L(PAH~8\mu m)/L(24~\mu m)\right)$ ratio color-coded by $\Sigma\left(TIR\right)$, the total infrared luminosity per unit area, which is a proxy to the star formation intensity at high metallicities. In Figure~\ref{fig:fit-subregions-SFR} the fit parameters for the subregions are plotted versus $\Sigma\left(TIR\right)$.
                                                                                                 
We see that the two branches in Figure~\ref{fig:PAH-24-TIR-SFR} are separated by $\Sigma\left(TIR\right)$. Indeed, most galaxies and galaxy subregions that have a low $\Sigma\left(TIR\right)$ tend to be on the lower branch, the $L(24~\mu m)/L(TIR)$ ratio is mostly independent of the $L(PAH~8\mu m)/L(24~\mu m)$ ratio. Conversely, whole galaxies and galaxy subregions that have a higher $\Sigma\left(TIR\right)$ are on the upper branch. The $L(24~\mu m)/L(TIR)$ ratio becomes dependent on the $L(PAH~8\mu m)/L(24~\mu m)$ ratio on the upper branch compared to the lower one. That is, for higher $\Sigma\left(TIR\right)$, $L(24)$ increases as the infrared spectral energy distribution is hotter. This dichotomy could be due to the ``loss'' of infrared in low metallicity objects. However, most importantly we note that for galaxies that have a $L(PAH~8\mu m)/L(24~\mu m)$ ratio typically lower than $-0.5$~dex, the $L(24~\mu m)/L(TIR)$ ratio depends almost exclusively on $\Sigma\left(TIR\right)$. Galaxies that have a $L(PAH~8\mu m)/L(24~\mu m)$ ratio typically lower than $-0.5$~dex happen to have a low metallicity and as such are expected to be predominantly on the lower branch. However, the weakness or even the lack of PAH emission in these galaxies implies that the $L(24~\mu m)/L(TIR)$ ratio is very sensitive to $\Sigma\left(TIR\right)$ at low metallicities.

Figure~\ref{fig:fit-subregions-SFR}, made using the same method as for Figure~\ref{fig:fit-subregions}, confirms quantitatively the results described above. At low $\Sigma\left(TIR\right)$ the slope is shallow but increases to reach a local maximum around $\log \Sigma\left(TIR\right)\simeq34.3$ before decreasing slightly. We note that unsurprisingly the behavior is globally similar to the one determined for the oxygen abundance. Indeed, lower metallicity galaxies are more transparent as they contain less dust and therefore the oxygen abundance and $\Sigma\left(TIR\right)$ are partly correlated.

The observed trend also accounts for the relation determined by \cite{calzetti2005a} (yellow line in Figure~\ref{fig:PAH-24-TIR-SFR}) since those authors have fitted star-forming regions in M51 removing the contribution to TIR from the dust heating evolved populations, maximizing the effect of the star formation intensity.

\subsubsection{Quality of the correlations}

An important point to note is that the scatter around the best fit tends to be larger for whole galaxies compared to subregions. This is likely due to the diversity in the samples. It clearly shows that those samples present an important variety of infrared properties \citep[e.g.][]{dale2009a}. 

We also estimate the TIR luminosity per unit area from the 24~$\mu$m band, that has been shown to be an accurate SFR estimator by \cite{rieke2009a}. This is particularly useful in case no other mid- or far-infrared wavelength is available. The scatter around the best fit is 0.087~dex, yielding an estimate of the TIR luminosity with an accuracy slightly over 20\%. Unsurprisingly, both luminosities per unit area are an increasing function of the oxygen abundance (Pearson correlation coefficient $r=0.45$ for the 24~$\mu$m emission and $r=0.53$ for the total infrared emission). The slope of the correlation -- 0.842, shallower than what was found by \cite{rieke2009a} -- shows that the 24~$\mu$m by itself is not a linear tracer of the TIR luminosity. Indeed, at higher luminosities, an increasing fraction of the infrared emission can be accounted for by the hot dust traced by the 24~$\mu$m band.

\subsubsection{Summary}

It appears that both the metallicity and $\Sigma\left(TIR\right)$ play a role in the determination of the TIR emission from the 8~$\mu$m and 24~$\mu$m bands. While those two parameters are intertwined (a low metallicity galaxy tends to be more transparent and hence have a lower TIR luminosity), they are the dominant parameters in different regimes.

For $L(PAH~8\mu m)/L(24~\mu m)<-0.5$~dex, $L(PAH~8\mu m)/L(24~\mu m)$ depends only on $\Sigma\left(TIR\right)$.

For $L(PAH~8\mu m)/L(24~\mu m)>-0.5$~dex two branches can be seen. For $\Sigma\left(TIR\right)\lesssim34.3$~dex, $L(24~\mu m)/L(TIR)$ is very weakly dependent on $L(PAH~8\mu m)/L(24~\mu m)$, higher metallicity galaxies and subregions tend to have a higher $L(PAH~8\mu m)/L(24~\mu m)$ ratio this being due to the increasingly stronger PAH emission with the metallicity up to a ratio of $\sim0.7$~dex. For higher metallicity galaxies and subregions, an upper branch is described, the $L(24~\mu m)/L(TIR)$ ratio increasing with $\Sigma\left(TIR\right)$ while the $L(PAH~8\mu m)/L(24~\mu m)$ ratio decreases, probably due to a combination of both the heating of the very small grains which strongly increases the 24~$\mu$m luminosity and possible a destruction of PAH in intense radiation fields. In other words, the main parameters are the metallicity which drives the strength of the PAH depending on the presence of dust carriers, and the star formation intensity which drives the temperature of the dust controlling the luminosity at 24~$\mu$m.

\subsection{TIR estimates that include the 70~$\mu$m band}

\begin{figure*}[!htbp]
\includegraphics[width=\columnwidth]{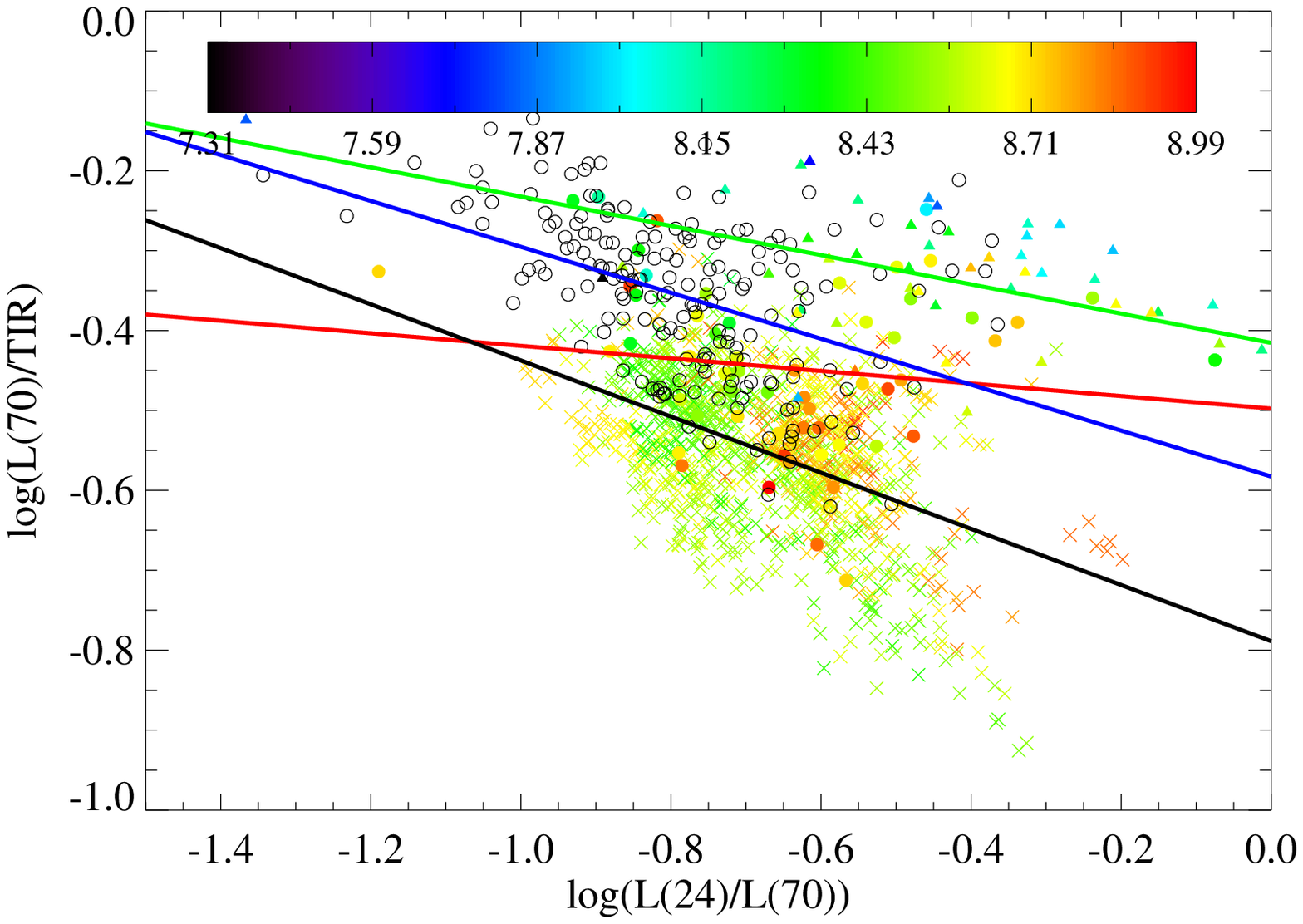}
\includegraphics[width=\columnwidth]{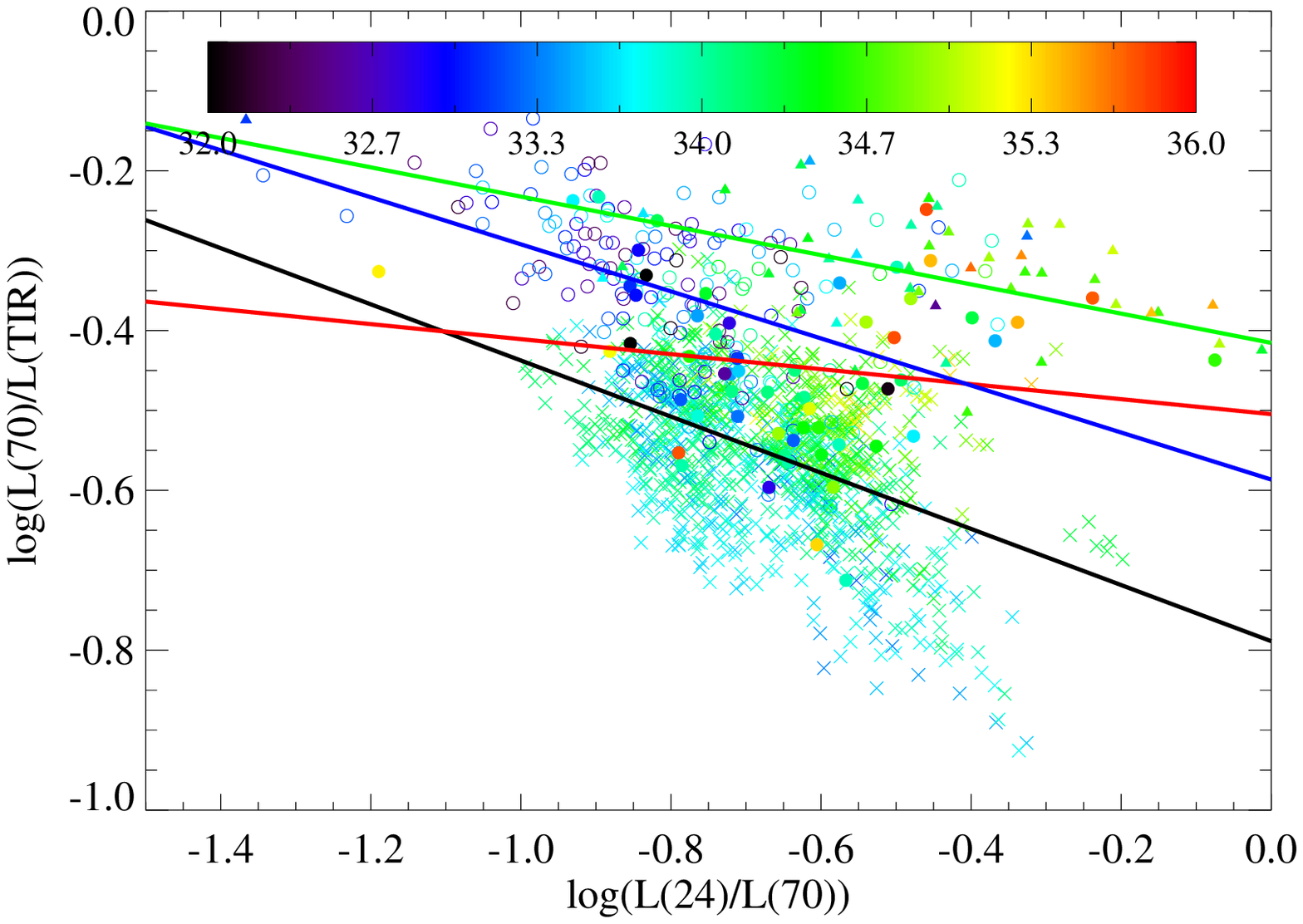}
\caption{$\log \left(L(70~\mu m)/L(TIR)\right)$ versus $\log \left(L(24~\mu m)/L(70~\mu m)\right)$. The symbols and the color-coding of the lines are the same as in Figure~\ref{fig:PAH-24-TIR}. The color codes the oxygen abundance (left) and $\Sigma\left(TIR\right)$ (right). \label{fig:70}}
\end{figure*}

Here we demonstrate how to estimate the TIR luminosity combining the 70~$\mu$m and 24~$\mu$m bands. The fits are presented in Figure~\ref{fig:70} and the numerical relations in Tables~\ref{tab:relations}, \ref{tab:relations-sigma} and \ref{tab:relations-sigma-oxy}.

We see that $L(70~\mu m)/L(TIR)$ depends significantly on the metallicity: low metallicity regions tend to have a high $L(70~\mu m)/L(TIR)$ ratio while higher metallicity regions tend to have a lower $L(70~\mu m)/L(TIR)$ ratio. Indeed, low metallicity galaxies tend to have a bluer $L(70~\mu m)/L(160~\mu m)$ color. This may demonstrate that the color temperature of the ``cool'' ($<$30~K) dust increases as metallicity decreases. This effect has by shown by \cite{helou1986a,engelbracht2008a} for instance. This could occur because the lower metallicity results in lower dust extinction, so a greater fraction of the dust that is present in low-metallicity systems is heated by strong radiation fields. In contrast, dust extinction in high metallicity systems would tend to make the radiation field appear less intense through most of the ISM. Consequently, most of the dust in high metallicity systems would be heated by a weaker radiation field and would appear cooler than the dust in low metallicity systems. The dependence of the fit parameters on $\Sigma\left(TIR\right)$ is weak. A similar result has been found by Calzetti et al. (2010, submitted).

In the relation between $L(24~\mu m)/L(70~\mu m)$ and $L(70~\mu m)/L(TIR)$, the integrated galaxy colors exhibit a shallower slope. Interestingly, about 31\% of LVL and E08 galaxies have $L(70~\mu m)/L(TIR)$ values greater than $-0.3$~dex but only 12\% of the SINGS galaxies and a mere 0.1\% of the galaxy subregions have $L(70~\mu m)/L(TIR)$ values that are this high.

The TIR luminosity per unit area is well correlated with the 70~$\mu$m one, and the scatter around the best fit is slightly smaller than when using the 24~$\mu$m band only (0.082 vs 0.087). This is due to the fact that large dust grains appear warmer in low metallicity galaxies. The slope -- 0.902 -- slightly under 1 makes the 70~$\mu$m is sub-linear tracer of the TIR luminosity which traces the star formation rate for galaxy subregions but nearly linear for galaxy samples. See Calzetti et al. (2010, submitted) and Li et al. (2010, in preparation) for the use of the 70~$\mu$m band as a star formation tracer. Taking into account the oxygen abundance in Table~\ref{tab:relations-sigma-oxy} improves the precision of the relation. Indeed, as stated earlier, the dust temperature is a function of the metallicity. Providing the oxygen abundance permits to indirectly take into account the dust temperature and therefore correct the estimation of the total infrared emission from the 70~$\mu$m band only.

\subsection{TIR estimates that include the 160~$\mu$m band}

\begin{figure*}[!htbp]
\includegraphics[width=\columnwidth]{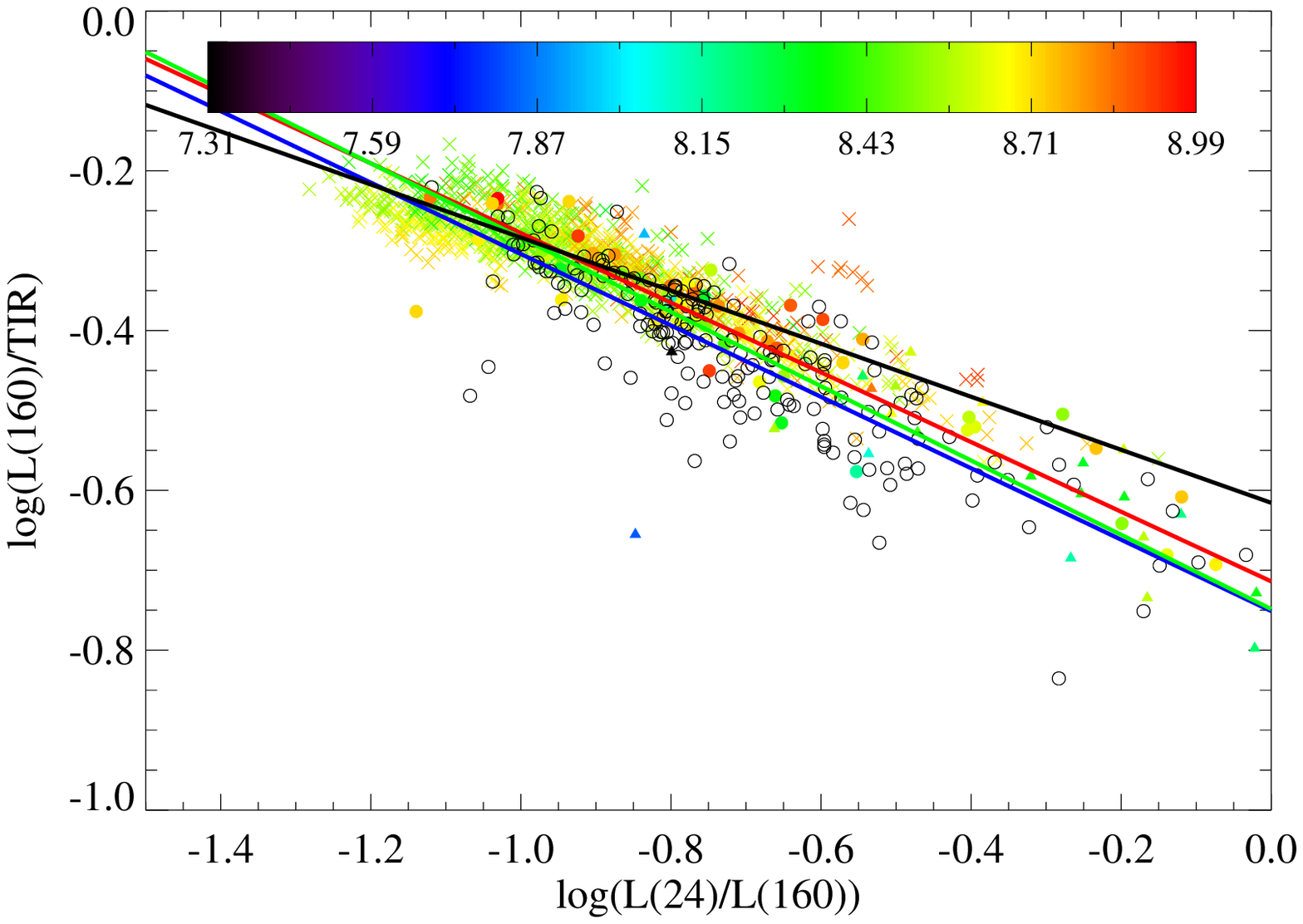}
\includegraphics[width=\columnwidth]{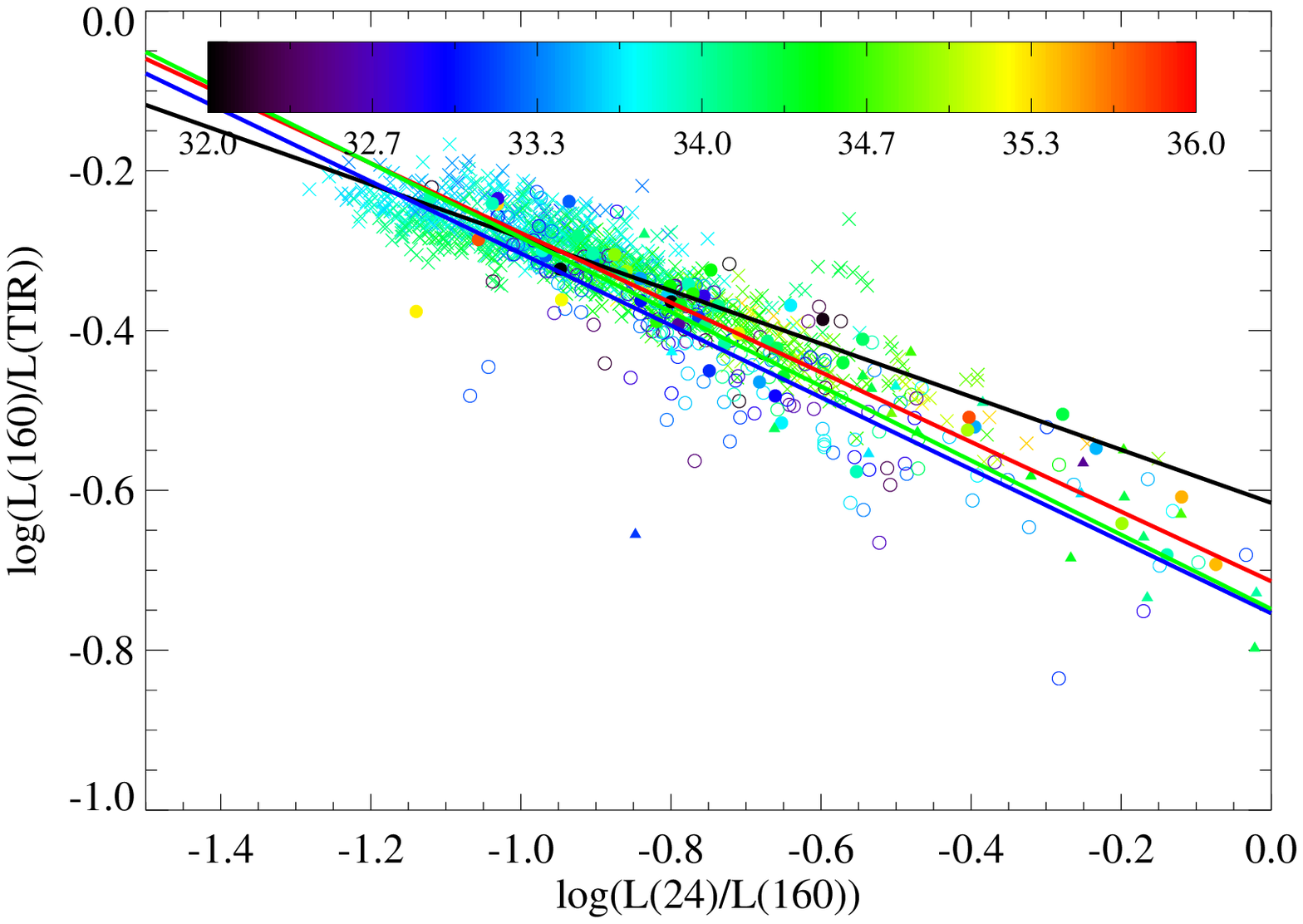}
\caption{$\log \left(L(160~\mu m)/L(TIR)\right)$ versus $\log \left(L(24~\mu m)/L(160~\mu m)\right)$. The symbols and the color-coding of the lines are the same as in Figure~\ref{fig:PAH-24-TIR}. The color codes the oxygen abundance (left) and $\Sigma\left(TIR\right)$ (right).\label{fig:160-24}}
\end{figure*}

The 160~$\mu$m band probes the cold dust, which makes it a good proxy to the TIR luminosity in galaxies where it is the dominant source of infrared emission \citep[see for example][]{bendo2008a}. In Figure~\ref{fig:160-24}, we show the correlation between $\log \left(L(24~\mu m)/L(160\mu m)\right)$ and $\log \left(L(160\mu m)/L(TIR)\right)$. In Table \ref{tab:relations} we see that the scatter around the best fit is smaller than for the 24~$\mu$m and the 70~$\mu$m.

Interestingly, neither the metallicity nor $\Sigma\left(TIR\right)$ have a significant effect on the parameters of the fit between $L(24~\mu m)/L(160\mu m)$ and $L(160\mu m)/L(TIR)$ as we can see in Figure~\ref{fig:160-24}. This proves to be an advantage when the metallicity of the galaxy is not known or very uncertain. The 160~$\mu$m luminosity per unit area is tightly correlated with the TIR luminosity per unit area. Unsurprisingly the slope is slightly superlinear (1.107) and the scatter around the best fit is small (0.049 dex) making the 160~$\mu$m the best MIPS {\em Spitzer} band to trace the TIR emission along with the 70~$\mu$m band. The scatter around the best fit is significantly larger for the SINGS, LVL and E08 samples however the slope is almost completely linear, especially in the case of the SINGS and LVL samples. Once again, this is likely due to the diversity of the populations constituting the samples. Taking into account the oxygen abundance slightly reduces the scatter when estimating the TIR emission from the 160~$\mu$m one. As for the 70~$\mu$m band the reason is that the metallicity gives an indication of the possible dust temperature.

\begin{figure*}[!htbp]
\includegraphics[width=\columnwidth]{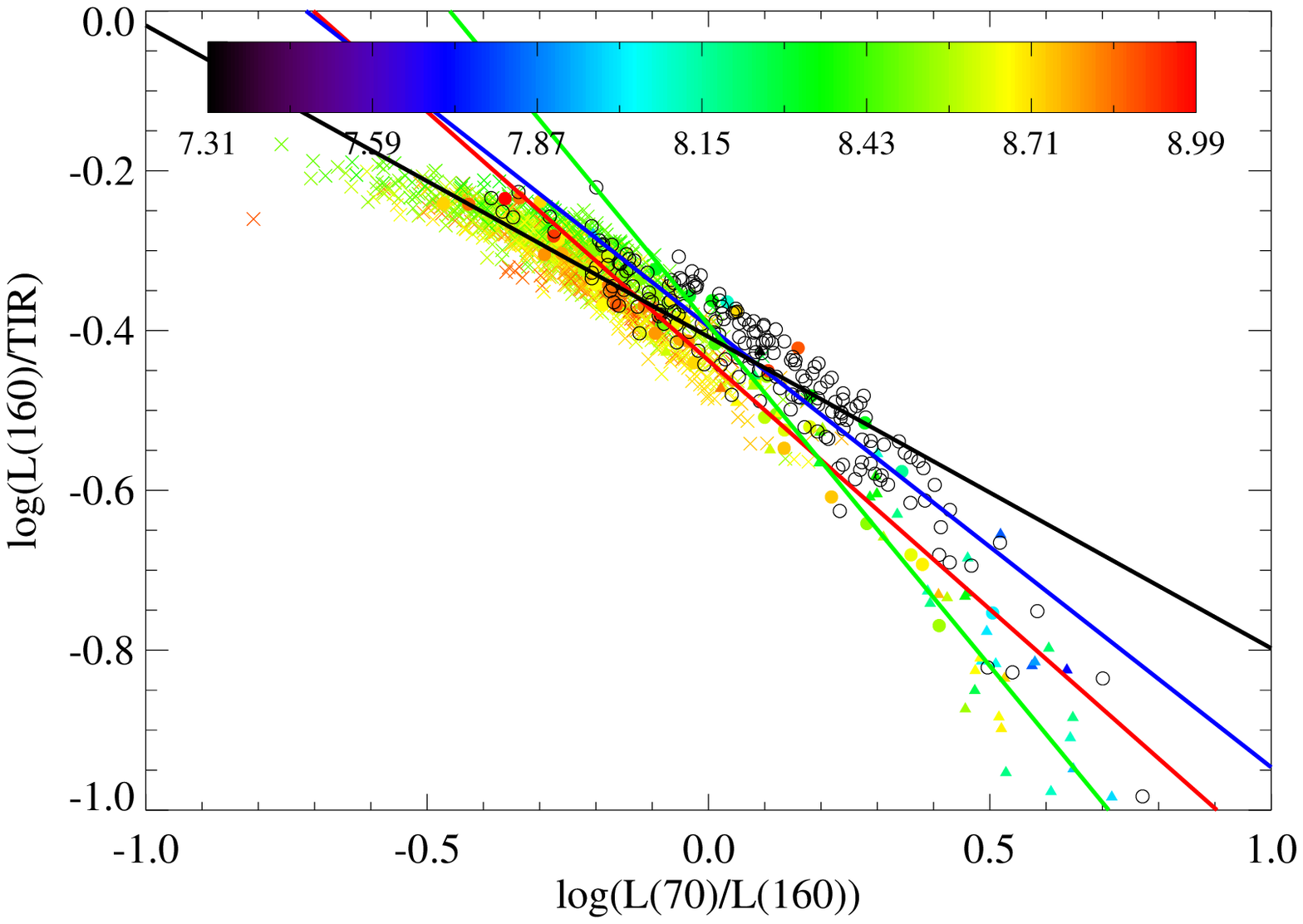}
\includegraphics[width=\columnwidth]{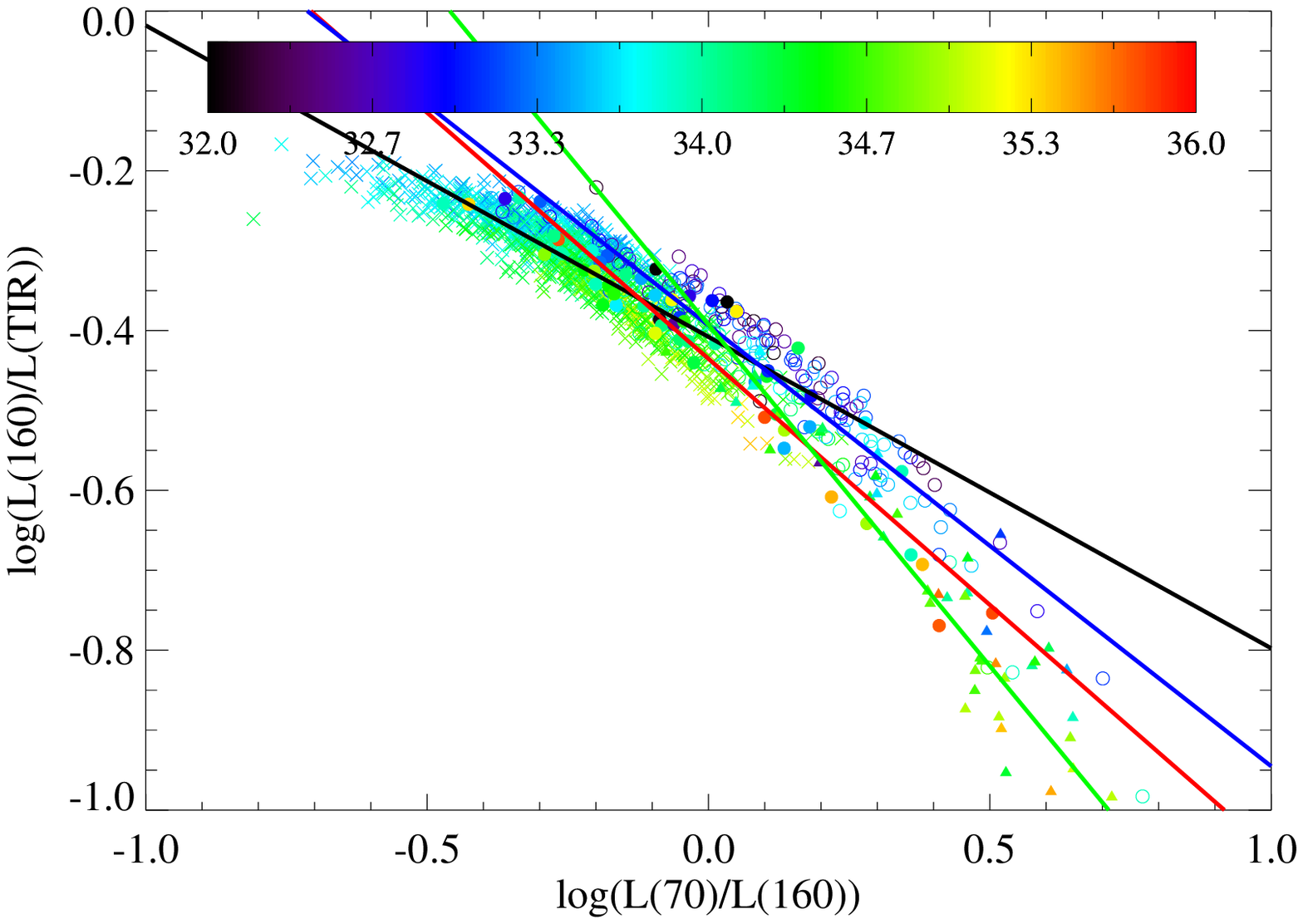}
\caption{$\log \left(L(70~\mu m)/L(160~\mu m)\right)$ versus $\log \left(L(160~\mu m)/L(TIR)\right)$. The symbols and the color-coding of the lines are the same as in Figure~\ref{fig:PAH-24-TIR}. The color codes the oxygen abundance (left) and $\Sigma\left(TIR\right)$ (right).\label{fig:160-70}}
\end{figure*}

As the 70~$\mu$m and 160~$\mu$m are major contributors to the TIR \citep[][Fig.~11c, f]{dale2009a}, their temperature should be a good indicator of the TIR emission. In Figure~\ref{fig:160-70}, we show the correlation between $\log \left(L(70~\mu m)/L(160\mu m)\right)$ and $\log \left(L(160\mu m)/L(TIR)\right)$.
We observe that there is a tight non-linear relation between the temperature of the dust as traced by the $L(70~\mu m)/L(160\mu m)$ ratio. At low $L(70~\mu m)/L(160\mu m)$, which indicates that the dust is very cold, the bulk of the TIR is accounted for by the 160~$\mu$m. A larger fraction of the total infrared emission is accounted for by the 70~$\mu$m band as the dust temperature increases. It appears that unlike the estimation from the 8~$\mu$m and 24~$\mu$m only for instance, the parameters of the correlation do not seem to depend on the the metallicity or $\Sigma\left(TIR\right)$ in any significant way.

\section{Discussion}
\label{sec:discussion}

The relations derived in this paper are dependent on the measurement of the total infrared luminosities and oxygen abundances.

\begin{figure*}[!htbp]
\includegraphics[width=\columnwidth]{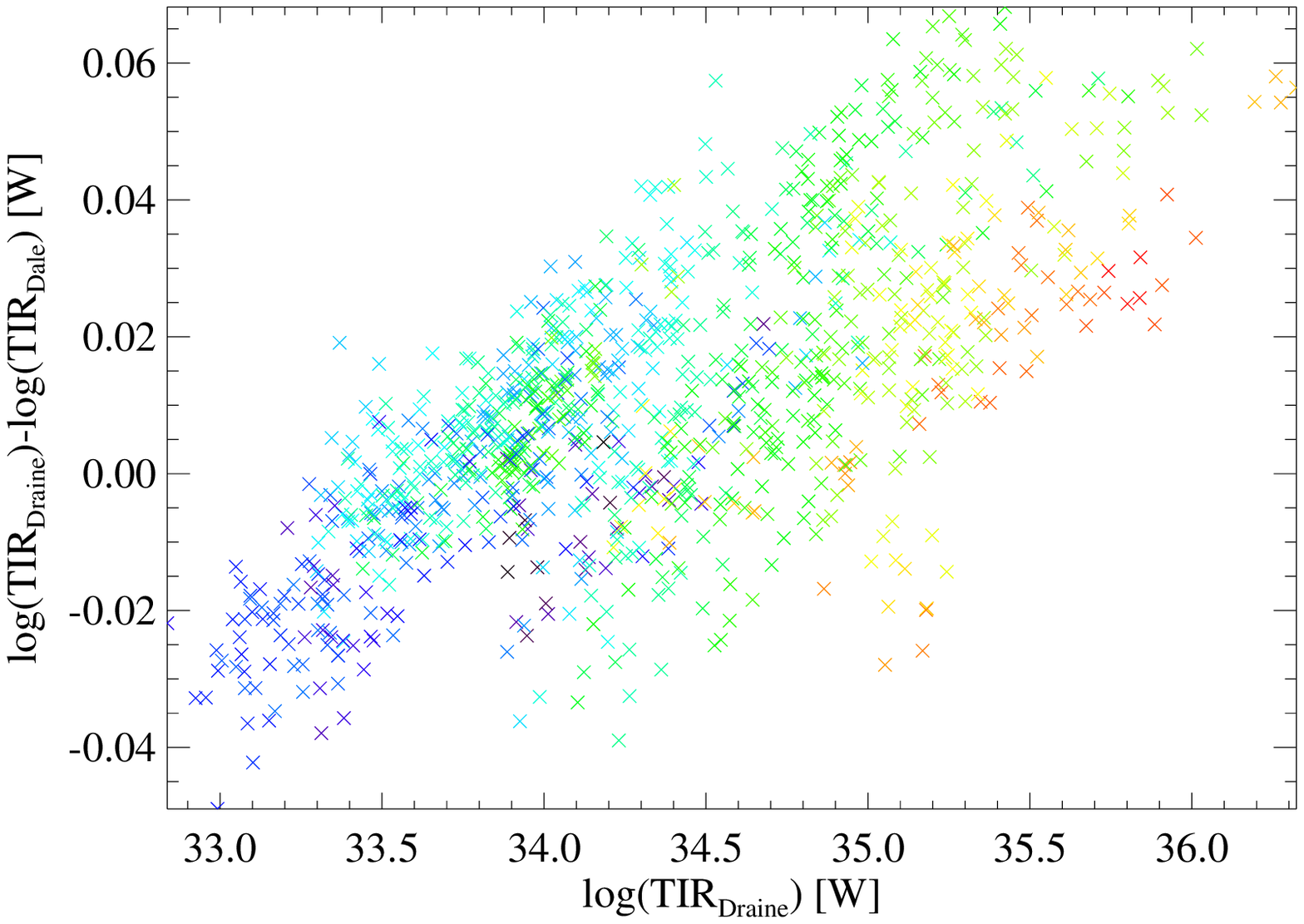}
\includegraphics[width=\columnwidth]{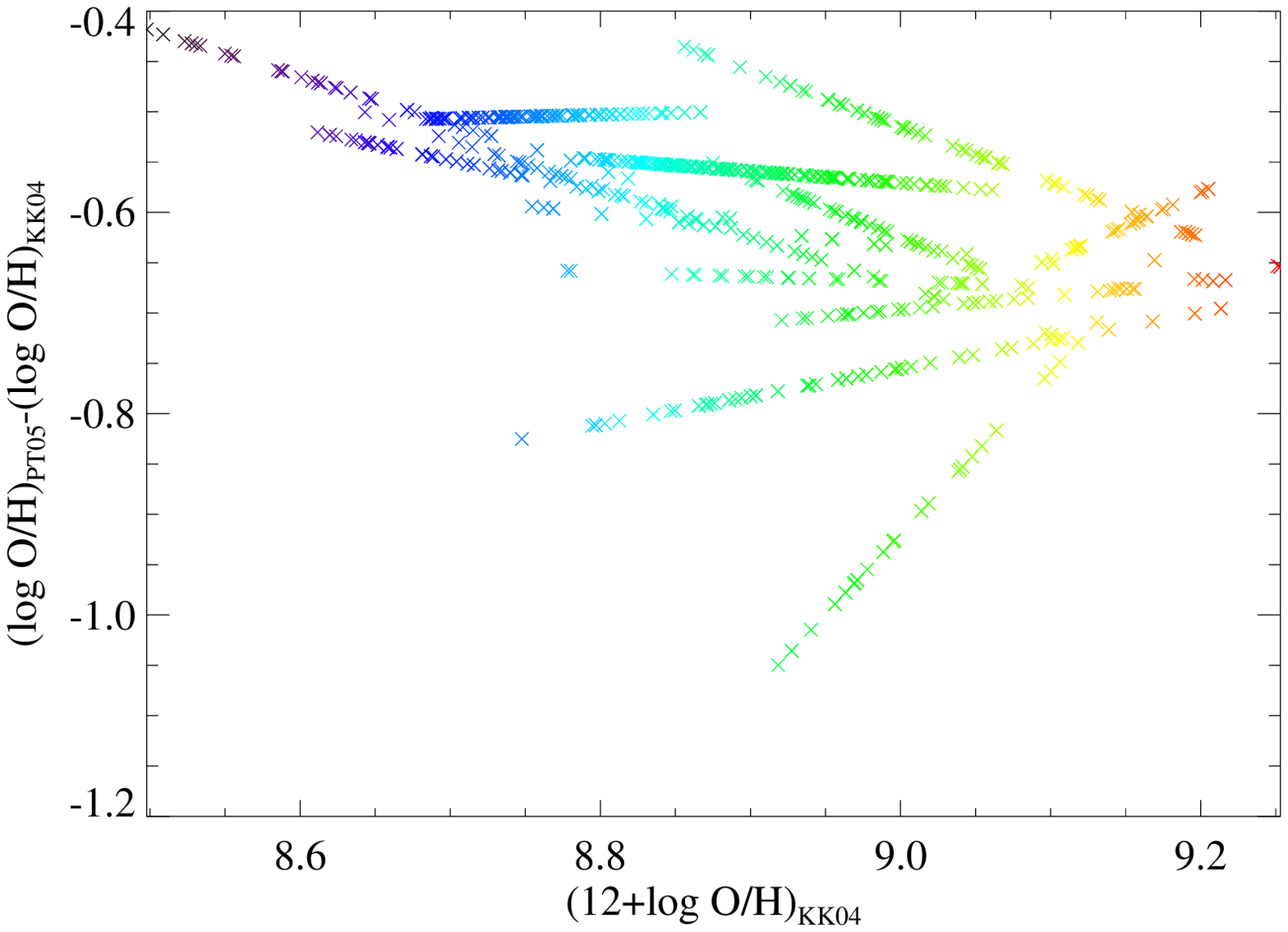}
\caption{Left: difference of the TIR luminosity estimated by the \cite{dale2001a} and \cite{draine2007a} relations versus the luminosity estimated from the \cite{draine2007a} relation. Right: difference between the oxygen abundances derived from the PT05 and the KK04 methods versus the latter one. The symbols and the colors are the same as in Figure~\ref{fig:PAH-24-TIR}.\label{fig:metal-draine-dale}}
\end{figure*}

We have estimated the TIR luminosity using equation \ref{eqn:draine} from \cite{draine2007a}. In the context of this study, equation 4 from \cite{dale2002a} yields very similar results. For all galaxy subregions, $\left<\log(TIR_{Draine}/TIR_{Dale})\right>=0.01\pm0.02$ as can be seen in the left panel of Figure~\ref{fig:metal-draine-dale}. The mean offset is similar for integrated SINGS and LVL galaxies data but slightly higher for the E08 sample: $\left<\log(TIR_{Draine}/TIR_{Dale})\right>=0.03\pm0.03$. The difference in the estimate of the TIR luminosity is small compared to the internal scatter within the subregions and the integrated galaxies which spans about 1 dex in $L(24~\mu m)/L(TIR)$. Therefore it should not affect the results significantly.

The method used to calculate the oxygen abundance is another source of uncertainty. The oxygen abundance estimates between the PT05 and KK04 are offset by $0.60\pm0.10$ dex. The shapes of the slope and y-intercept curves versus the oxygen abundance are not identical, similarly to what was done in Figure~\ref{fig:fit-subregions}. There are two reasons to this. First of all, the two estimators are not in linear relation. Then, the abundance gradients derived by Moustakas et al. (2010, in preparation) are different for the two estimators as can be seen in the right panel of Figure~\ref{fig:metal-draine-dale}. The consequence is that each subregion has a different oxygen abundance offset between the two estimators. The difference in the fit parameters between the two oxygen abundance estimators for a given observation provides the typical uncertainties due to the abundance estimators. However, the precise evaluation of the uncertainties would require resolved spectroscopic observations at a $\sim6$\arcsec\ resolution. If we take the difference of the extrema of the fit parameters using the KK04 and PT05 estimators, the typical uncertainty on the slope can be evaluated to about $\pm0.2$ and is the main uncertainty in this study. Even though the uncertainty is important, it is significantly smaller than the range spanned by the slope, $\sim1.2$ for the PT05 method and $~0.8$ for the KK04 one.

\begin{figure}[!ht]
\includegraphics[width=\columnwidth]{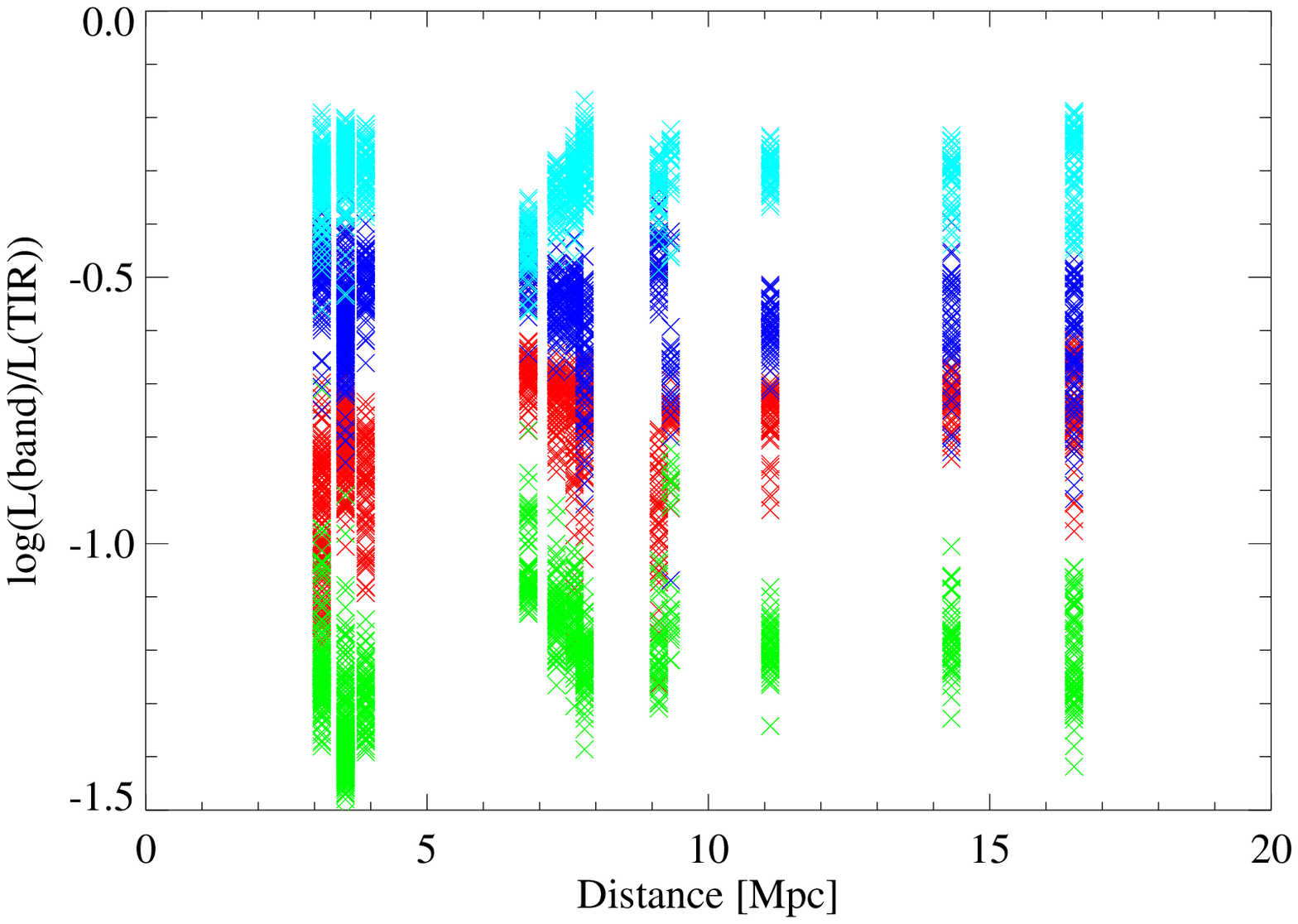}
\caption{Ratio of the luminosity in the PAH 8~$\mu$m (red), 24~$\mu$m (green), 70~$\mu$m (blue), 160~$\mu$m (cyan) bands for each pixel as a function of the distance of the galaxy.\label{fig:test-res-effect}}
\end{figure}

Even if the relations derived here are distance independent, the physical scale encompassed by a pixel ranges from 0.7~kpc to 3.6~kpc in galaxy subregions. Even for the closest galaxy, each pixel encompasses several star forming regions and the ISM of the galaxy. However, for the more distant galaxies the mixing of the two components increases with the physical scale encompassed. To check whether this effect could induce a distance dependent bias in our analysis, we plot in Figure~\ref{fig:test-res-effect} the ratio of the luminosity in each infrared band to the total infrared luminosity for each pixel, as a function of the distance of the galaxy.
We see that there is no obvious trend with the distance.

\section{Conclusions}
\label{sec:conclusion}

Using data for spatially resolved subregions within 13 face-on spiral galaxies as well as integrated luminosities for the SINGS, LVL, and E08 galaxies, we have derived new relations to estimate the total infrared luminosity based on using only one or two {\em Spitzer} bands, particularly the 8.0~$\mu$m and 24~$\mu$m bands. Relations incorporating 8.0~$\mu$m data vary significantly with oxygen abundance and especially with $\Sigma\left(TIR\right)$. However, TIR estimates that do not include 8~$\mu$m data are less dependent on oxygen abundances. In particular, the relations between the TIR emission and the 70~$\mu$m or 160~$\mu$m bands are relatively independent of oxygen abundances compared to the 8~$\mu$m and 24~$\mu$m ones.

\appendix

\section{Recipes to calculate the TIR luminosity}

We provide here a few general guidelines to use the relations derived in this paper:
\begin{itemize}
 \item If available, the 70~$\mu$m and 160~$\mu$m bands should be used to estimate the total infrared emission as they provide the most accurate result,
 \item If only one band is available, the relations provided in Tables~\ref{tab:relations-sigma} and \ref{tab:relations-sigma-oxy} should be used provided the distance is available and the target galaxy is resolved,
 \item If only one band is available, and an estimation of the oxygen abundance is available, the relations provided in Table~\ref{tab:relations-sigma-oxy} should be used,
 \item If only the 8~$\mu$m and 24~$\mu$m bands are available, both should be used to determine the TIR emission, in particular when the luminosity per unit area is not available.
\end{itemize}

We list in Table~\ref{tab:recipes} the most appropriate formulas to be used to derive the total TIR luminosity of a galaxy, based on our results parametrized as a function of available information: IR data (single or multiple bands), metallicity, and expected $\Sigma\left(TIR\right)$.

\begin{deluxetable*}{ccccccc}
\tablecolumns{7} \tablewidth{0pc} \tablecaption{TIR
 estimations recipes\label{tab:recipes}}
\tablehead{\colhead{8}&\colhead{24}&\colhead{70}&\colhead{160}&\colhead{Z?}&\colhead{$\Sigma\left(TIR\right)$?}&\colhead{Recipe}}
\startdata
x&&&&Low/Mid&High&E08\\
x&&&&High&Any/Unknown&Subregions\\
x&&&&Any/Unknown&Any/Unknown&LVL/SINGS\\
x&x&&&Low/Mid&Low&LVL\\
x&x&&&Low/Mid&High&E08\\
x&x&&&High&Any/Unknown&Subregions\\
x&x&&&Unknown&Unknown&SINGS\\
&x&&&High&Any/Unknown&Subregions\\
&x&&&Low/Mid&High&E08\\
&x&&&Any/Unknown&Any/Unknown&LVL/SINGS\\
&x&x&&High&Any/Unknown&Subregions\\
&x&x&&Low/Mid&High&E08\\
&x&x&&Any&Any&LVL/SINGS\\
&x&&x&High&Any/Unknown&Subregions\\
&x&&x&Any/Unknown&Any/Unknown&LVL/SINGS/E08\\
&&x&&High&Any/Unknown&Subregions\\
&&x&&Any/Unknown&Any/Unknown&LVL/SINGS/E08\\
&&x&x&Any/Unknown&Any/Unknown&Global\\
&&&x&Low/Mid&High&E08\\
&&&x&Any/Unknown&Any/Unknown&LVL/SINGS/Subregions
\enddata
\tablecomments{The low, mid and high ranges are only indicative and are defined relatively to the range spanned by the samples.}
\end{deluxetable*}

\acknowledgments

We thank the anonymous referee for very useful comments that have helped improving and clarifying the paper. This work has been supported by NASA ADP grant NNX07AN90G. MB thanks Alexey Vikhlinin for his technical help during the redaction of the manuscript.

{\it Facilities:} \facility{Spitzer}
\bibliographystyle{aa}
\bibliography{article}

\end{document}